\RequirePackage[l2tabu,orthodox]{nag}
\documentclass[aps,prl,twocolumn,superscriptaddress,showpacs,amsmath,amsfonts,amssymb,floatfix,nofootinbib]{revtex4-1}

\usepackage{bm}
\usepackage{color}
\usepackage{float}
\usepackage{graphicx} \graphicspath{{./Images/}}
\usepackage{hyperref}
\usepackage[capitalise]{cleveref}
\usepackage{xstring} \noexpandarg
\usepackage{xparse}

\usepackage{acro}

\DeclareAcronym{TDSE}  { short = TDSE, long = time-dependent Schr\"odinger equation }
\DeclareAcronym{TDDFT} { short = TDDFT, long = time-dependent density-functional theory }
\DeclareAcronym{KS} { short = KS, long = Kohn-Sham }
\DeclareAcronym{Mxc} { short = Mxc, long = mean-field exchange-correlation }
\DeclareAcronym{Mx} { short = Mx, long = mean-field exchange }
\DeclareAcronym{TDOEP} { short = TDOEP, long = time-dependent optimized-effective potential }

\begin{document}

\newcommand \diff {\mathrm{d}}
\newcommand \imagi {\mathrm{i}}
\newcommand \br {\mathbf{r}}
\newcommand \bz {\mathbf{z}}
\newcommand \blambda {\bm{\lambda}}
\newcommand \bj {\mathbf{j}}
\newcommand \bF {\mathbf{F}}

\NewDocumentCommand \bra{r<|}{\langle #1 \vert}
\NewDocumentCommand \ket{r|>}{\vert #1 \rangle}
\NewDocumentCommand \braket{r<>}{\langle \StrSubstitute{#1}{|}{\vert } \rangle}
\RenewDocumentCommand \outer{r|>r<|}{\vert #1 \rangle \langle #2 \vert}
\NewDocumentCommand \Bra{r<|}{\left\langle #1 \right\vert}
\NewDocumentCommand \Ket{r|>}{\left\vert #1 \right\rangle}
\NewDocumentCommand \Braket{r<>}{\left\langle \StrSubstitute{#1}{|}{\middle\vert } \right\rangle}
\NewDocumentCommand \Outer{r|>r<|}{\left\vert #1 \middle\rangle \middle\langle #2 \right\vert}


\title{Dressed-Orbital Approach to Cavity Quantum Electrodynamics and Beyond}

\author{Soeren Ersbak Bang Nielsen}
\email[Electronic address:\;]{soerenersbak@hotmail.com}
\author{Christian Sch\"afer}
\email[Electronic address:\;]{christian.schaefer.physics@gmail.com}
\author{Michael Ruggenthaler}
\email[Electronic address:\;]{michael.ruggenthaler@mpsd.mpg.de}
\affiliation{Max Planck Institute for the Structure and Dynamics of Matter and Center for Free-Electron Laser Science, Luruper Chaussee 149, 22761 Hamburg, Germany}
\author{Angel Rubio}
\email[Electronic address:\;]{angel.rubio@mpsd.mpg.de}
\affiliation{Max Planck Institute for the Structure and Dynamics of Matter and Center for Free-Electron Laser Science, Luruper Chaussee 149, 22761 Hamburg, Germany}
\affiliation{Center for Computational Quantum Physics (CCQ), The Flatiron Institute, 162 Fifth avenue, New York NY 10010}

\date{\today}

\begin{abstract}
	We present a novel representation of coupled matter-photon systems that allows the application of many-body methods developed for purely fermionic systems.
	We do so by rewriting the original coupled light-matter problem in a higher-dimensional configuration space and then use photon-dressed orbitals as a basis to expand the thus "fermionized" coupled system.
	As an application we present a dressed \acl*{TDDFT} approach.
	The resulting dressed \acl*{KS} scheme allows for straightforward non-adiabatic approximations to the unknown exchange-correlation potential that explicitly includes correlations.
	We illustrate this for simple model systems placed inside a high-Q optical cavity, and show also result for observables such as the photon-field fluctuations that are hard to capture in standard matter-photon \acl*{KS}.
	We finally highlight that the dressed-orbital approach extends beyond the context of cavity quantum electrodynamics and can be applied to, e.g., van-der-Waals problems.
\end{abstract}
\maketitle


	In the last decade experiments at the interface between chemistry, material science and quantum optics have uncovered situations in which the strong interplay between the quantized electromagnetic field and the matter degrees of freedom lead to interesting physical phenomena and novel states of matter \cite{ebbesen2016, sukharev2017, flick2017a, ruggenthaler2018quantum}.
	For instance, chemical reactions can be modified by strongly coupling molecules to the vacuum field of an optical cavity \cite{hutchison2012, thomas2016}, strong exciton-photon coupling in light-harvesting complexes is expected to modify energy transfer \cite{coles2014}, the coupling to quantum matter can lead to attractive photons \cite{firstenberg2013} and the ultra-strong coupling to artificial atoms in circuit quantum electrodynamics extends far beyond the perturbative regime \cite{diaz2016}.
	Experimental results show that the emergence of hybrid light-matter states, so-called polaritonic states that strongly mix matter and photon degrees of freedom, is the reason for these interesting phenomena.
	While the theoretical understanding of these effects are mostly based on simplified Dicke-type models (several few-level systems coupled to one mode) \cite{Galego2016}, an accurate and unbiased description of the physical situation calls for calculations of the coupled matter-photon systems from \textit{first principles} \cite{ruggenthaler2018quantum, flick2017a, george2016, schafer2018ab}.
	An appealing such first-principles method is an extension of density-functional theory to coupled matter-photon systems \cite{ruggenthaler2011, tokatly2013, ruggenthaler2014, ruggenthaler2015}, which is called quantum-electrodynamical density-functional theory.
	By reformulating the fully coupled fermion-boson problem in a formally exact quantum-fluid description, where the charge current is coupled in a non-linear way to the Maxwell field, it avoids a solution in terms of the usually infeasible wave function.
	The major drawback of such density-functional approaches is that the internal force terms of the quantum fluids are only known implicitly (in terms of the wave function) and approximations have to be used in general.
	The most successful approximation strategy is the \ac{KS} scheme, where the local-force expressions of a non-interacting auxiliary system is used as a starting point to model the fully interacting case \cite{flick2015, pellegrini2015}.
	First calculations for real molecules coupled to photons in and out of equilibrium \cite{flick2018ab,flick2018light,flick2018} show the potential of first-principle calculations of coupled matter-photon systems.
	However, approximations based on the standard \ac{KS} scheme are hard to improve towards strongly-coupled systems, which in the context of coupled matter-photon systems promise interesting physical effects \cite{diaz2016,flick2017a,galego2017,schafer2019modification}.
	Alternative approximation schemes either rely on a different auxiliary system, such as the strictly-correlated electron system \cite{gorigiorgi2009} or go beyond the single Slater-determinant Ansatz \cite{fuks2016}.

	In this work we provide a completely different route to describe matter-photon systems from first principles by reformulating the physical problem in a space with auxiliary extra dimensions.
	This higher-dimensional reformulation reduces to the standard formulation in physical space in a straightforward manner, providing us with an ``holographic'' perspective of the original problem that allows us to work with explicitly-correlated/dressed higher-dimensional orbitals.
	Since this leads to a "fermionization" of the coupled fermion-boson system it allows us to employ fermionic many-body methods such as Greens' function techniques \cite{fetter2012,stefanucci2013} or density-functional methods.
	Here we exemplify the possibilities of this approach by applying it in the context of \ac{TDDFT} for the case of an electronic quantum system coupled to the photons of an optical cavity.
	For this dressed \ac{KS} scheme already the simplest approximations are non-adiabatic and include explicit correlations that can otherwise only be captured by advanced functionals for the standard \ac{KS} scheme.
	We show how simple approximations in terms of dressed (mixed matter-photon) orbitals capture the right Rabi-oscillation induced by the photon-matter coupling for a Rabi model and the spontaneous emission of a ``bare'' model helium that is brought inside an optical cavity.


	Let us consider the case of a general electronic system with frozen ions inside an optical cavity (the extension to include the nuclei as quantum particles is straightforward \cite{flick2017a, schafer2018ab}).
	As the spatial extension of the matter system is small compared to the wave length of the cavity modes, we can treat the matter-photon coupling in dipole approximation (atomic units are used throughout) \cite{grynberg2010, rokaj2017}.
	In this case the interacting Hamiltonian reads
\begin{align*}
\hat{H}(t) &= \sum_{k=1}^N \left[ -\tfrac{1}{2} \nabla_{\br_k}^2 + v(\br_k, t) \right] + \tfrac{1}{2} \sum_{k \neq l} w(\br_k,\br_l) \\
&\!\!\!\!\!+ \sum_{\alpha=1}^M \bigg[ - \tfrac{1}{2} \tfrac{\partial^2}{\partial p_\alpha^2} + \tfrac{1}{2} \Big( \omega_\alpha p_\alpha - \blambda_\alpha \cdot \sum_{k=1}^N \br_k \Big)^2 \!\! + \tfrac{\dot{J}_\alpha(t)}{ \omega_\alpha} p_\alpha \bigg] ,
\end{align*}
where the first line corresponds to the usual many-body Hamiltonian $\hat{T} + \hat{V}(t) + \hat{W}$ describing the uncoupled matter system of $N$ electrons interacting via the Coulomb potential $w(\br,\br')$ and moving in an external time dependent potential $v(\br,t)$.
	The second line describes $M$ photon modes with frequency $\omega_\alpha$ and polarization vectors $\blambda_\alpha$ coupled to the total dipole of the electronic system.
	Furthermore, the photon modes are allowed to couple to an external current $J_\alpha(t)$ \cite{tokatly2013, rokaj2017}.

	We first introduce auxiliary extra dimensions $(p_{\alpha,2}, ... , p_{\alpha,N})$ for each mode, to extend each $p_\alpha$ up to the electronic dimension $N$, and then consider the extended Hamiltonian $\hat{H}'(t) = \hat{H}(t) + \hat{H}_\mathrm{aux}$, with $\hat{H}_\mathrm{aux} = \sum_{\alpha=1}^M \hat{H}_{\mathrm{aux},\alpha}$ and $\hat{H}_{\mathrm{aux},\alpha} = \sum_{k=2}^N \left( -\tfrac{1}{2} \tfrac{\partial^2}{\partial p_{\alpha,k}^2} + \tfrac{\omega_\alpha^2}{2} p_{\alpha,k}^2 \right)$.
	We then change coordinates to $(q_{\alpha,1},...,q_{\alpha,N})$ for each mode, where $(q_{\alpha,1},...,q_{\alpha,N})$ is any coordinate set related to $(p_\alpha,p_{\alpha,2},...,p_{\alpha,N}$) by an orthogonal transformation (the specific choice does not matter) that further allows us to specify the photon-displacement coordinates $p_\alpha$ as
\begin{align} \label{CenterOfMass}
p_\alpha = \tfrac{1}{\sqrt{N}} \left( q_{\alpha,1} + ... + q_{\alpha,N} \right) .
\end{align}
	This is the inverse of a center-of-mass coordinate transformation \cite{watson1968} (see Supplemental Material Sec.~I for an explicit example).
	Introducing a $(3+M)$-dimensional dressed vector of space and auxiliary photon coordinates $\bz = (\br, q_1,...,q_M)$, i.e., each fermion is now a quasi-particle (polariton), we can then rewrite $\hat{H}'(t)$ as
\begin{align} \label{DressedHamiltonian}
\hat{H}'(t) &= \sum_{k=1}^N \left[ -\tfrac{1}{2} \nabla_{\bz_k}^2 + v'(\bz_k,t) \right] + \tfrac{1}{2} \sum_{k \neq l} w'(\bz_k,\bz_l) \\
&= \hat{T}' + \hat{V}'(t) + \hat{W}' . \nonumber
\end{align}
	Here \scalebox{0.86}[1]{$v'(\bz,t) \!=\! v(\br,t) + \sum_{\alpha=1}^M [ \tfrac{1}{2} \omega_\alpha^2 q_\alpha^2 - \tfrac{\omega_\alpha}{\sqrt{N}} q_{\alpha} (\blambda_\alpha \cdot \br) + \tfrac{1}{2} (\blambda_\alpha \cdot \br)^2$} $+ \tfrac{\dot{J}_\alpha(t) q_{\alpha}}{\sqrt{N} \omega_\alpha} ]$ and $w'(\bz,\bz') =  w(\br,\br') + \sum_{\alpha=1}^M [ -\tfrac{\omega_\alpha}{\sqrt{N}} q_{\alpha} (\blambda_\alpha \cdot \br')$ $- \tfrac{\omega_\alpha}{\sqrt{N}} q_{\alpha}' (\blambda_{\alpha} \cdot \br) + (\blambda_\alpha \cdot \br) (\blambda_\alpha \cdot \br')]$, i.e., $w'(\bz,\bz')$ represents "polariton-polariton" interactions.
	Here we used that for an orthogonal transformation $\sum_{k=1}^N \tfrac{\partial^2}{\partial p_{\alpha,k}^2} = \sum_{k=1}^N \tfrac{\partial^2}{\partial q_{\alpha,k}^2}$ and $\tfrac{\omega_\alpha^2}{2}(p_\alpha^2+p_{\alpha,2}^2+...+p_{\alpha,N}^2) = \tfrac{\omega_\alpha^2}{2}(q_{\alpha,1}^2+...+q_{\alpha,N}^2)$.
	Now, if the normalized $\Psi(t)$ solves the physical \ac{TDSE} $\imagi \tfrac{\partial}{\partial t} \Psi(t) = \hat{H}(t) \Psi(t)$, then
\begin{align*}
&\Psi'(\bz_1 \sigma_1, ..., \bz_N \sigma_N, t) = \\
&\Psi(\br_1 \sigma_1, ... , \br_N \sigma_N, p_1, ..., p_M, t) \chi(p_{1,2}, ..., p_{M, N},t) ,
\end{align*}
is the normalized solution of the extended \ac{TDSE} $\imagi \tfrac{\partial}{\partial t} \Psi'(t) = \hat{H}'(t) \Psi'(t)$ with $\chi(t)$ the state that evolves from an arbitrary, normalised initial state $\chi_0$ under $\hat{H}_\mathrm{aux}$.
	Since $\Psi(t)$ depends only on the coordinates defined in \cref{CenterOfMass}, it is invariant under exchange of $q_{\alpha,k}$ with $q_{\alpha,l}$.
	If we further pick $\chi_0$ symmetric with respect to exchange of $q_{\alpha,k}$ and $q_{\alpha,l}$, $\Psi'(t)$ is anti-symmetric under exchange of $\bz_k\sigma_k$ and $\bz_l\sigma_l$.
	It can therefore be expanded in Slater-determinants of $(3+M)$-dimensional dressed orbitals $\varphi'(\bz \sigma)$.
	These observations make the application of well-established many-body methods such as the \ac{KS} approach to \ac{TDDFT} possible.
	We refer to Supplemental Material Sec.~II for further details.%
\footnote{%
	Note that $\hat{H}'$ also has many eigenstates not in the form $\Psi'=\Psi\chi$, which are not physical.
	To compute only the physical eigenstates directly from $\hat{H}'$, one therefore still has to enforce the separation in $\Psi\chi$, $\br_k\sigma_k \leftrightarrow \br_l\sigma_l$ anti-symmetry and $q_{\alpha,k} \leftrightarrow q_{\alpha,l}$ symmetry.
	Other methods must also be adjusted to respect these properties.
}

	The \ac{KS} approach \cite{ruggenthaler2015ex} maps the interacting many-body problem of \cref{DressedHamiltonian} to a non-interacting auxiliary problem, i.e., $\hat{H}'_\mathrm{KS}(t) = \hat{T}' + \hat{V}'_\mathrm{KS}(t)$.
	This auxiliary dressed \ac{KS} system, usually given in terms of a Slater determinant $\Phi'(t)$ of dressed orbitals with spatial part $\varphi'_k(\bz,t)$, is enforced to generate the same (3+M)-dimensional expectation value
\begin{align*}
n'(\bz,t) &= \braket<\Psi'(t)|\hat{n}'(\bz)|\Psi'(t)> \\
&= N \sum_{\sigma_1,...,\sigma_N} \int \diff^{(3+M)(N-1)} z |\Psi'(\bz \sigma_1,...,\bz_N \sigma_N)|^2,
\end{align*}
of the dressed density operator $\hat{n}'(\bz) = \sum_{k=1}^N \delta(\bz - \bz_k)$.
	To ensure this a \ac{Mxc} potential $v'_\mathrm{Mxc}$ is introduced that mimics the interactions.
	This leads to non-linear dressed single-particle equations
\begin{align*}
\imagi \tfrac{\partial}{\partial t} \varphi_k'(\bz,t) =
\left[ -\tfrac{1}{2} \nabla_\bz^2 + v'(\bz,t) + v_\mathrm{Mxc}'(\bz,t) \right] \varphi_k'(\bz,t) ,
\end{align*}
where the exact dressed density is given by $n'(\bz,t) = \sum_{k=1}^N |\varphi_k'(\bz,t)|^2$.
	Similarly we can of course also set up a dressed ground-state density-functional theory \cite{gross2013}.
	By construction the dressed density reduces to the exact electron density via $n(\br,t) = \int \diff^M q \, n'(\bz,t)$ and to the exact expectation value of $p_\alpha(t) = \int \diff^{(3+M)} z \, \tfrac{q_\alpha}{\sqrt{N}} n'(\bz,t)$.

	In order for the dressed \ac{KS} approach to be practical we need two things.
	The first is to be able to handle the dimensionality of the dressed orbitals.
	In most situations of cavity quantum electrodynamics only one mode is important (i.e., strongly correlated with the matter), and the application of this approach is only one dimension more expensive than standard \ac{KS} (i.e., \cite{tokatly2013}).
	This renders dressed \ac{KS} especially appealing.
	On the other hand, if more than one cavity mode matters, we can adopt a reduction of the basis set similar to the calculations done in Ref. \cite{flick2017a} for a multi-mode cavity, where only up to a few-photon states are considered.
	In the case that many photons are involved, a simple mean-field treatment within the standard \ac{KS} approach becomes accurate again \cite{flick2015} and a dressed approach becomes less attractive.
	The other thing we need for the dressed \ac{KS} approach to be practical is an approximation for the \ac{Mxc} potential.
	In this regard it is interesting to compare the equations of motion for the physical and the dressed \ac{KS} systems (for the derivations, and a more complete analysis, we refer to Supplemental Material Sec.~III).
	The physical equation of motion for the density \cite{tokatly2005,tokatly2013}, obtained by Heisenberg's equations, is given by
\begin{align*}
\tfrac{\partial^2}{\partial t^2} n(\br,t) &= \nabla_\br \cdot \left[ n(\br,t) \nabla_\br v(\br,t) \right] \\
& - \nabla_\br \cdot [\mathbf{Q}(\br,t) + \bF_\mathrm{dip}(\br,t) + \bF_\mathrm{lin}(\br,t)] ,
\end{align*}
where $\mathbf{Q}(\br,t) = \imagi \braket<\Psi(t)|[\hat{T}+\hat{W},\hat{\bj}(\br)]|\Psi(t)>$ is the physical momentum-stress and interaction-stress forces.
	In turn, $\hat{\bj}(\br) = \tfrac{1}{2\imagi} \sum_{k=1}^N \, [\delta(\br-\br_k) \overrightarrow\nabla_{\br_k} - \overleftarrow\nabla_{\br_k} \delta(\br-\br_k)]$ is the physical current operator.
	Further,
\begin{align*}
\bF_\mathrm{dip}(\br,t) &= - \sum_{\alpha=1}^M \blambda_\alpha \braket<\Psi(t)| (\blambda_\alpha \cdot \sum_{k=1}^N \br_k ) \hat{n}(\br) |\Psi(t)> = \\
&\hspace{-14mm} - \sum_{\alpha=1}^M \blambda_\alpha n(\br t) (\blambda_\alpha \cdot \br)
- 2 \sum_{\alpha=1}^M \blambda_\alpha \int \diff \br' \rho_2(\br,\br',t) (\blambda_\alpha \cdot \br')  , \\
\bF_\mathrm{lin}(\br,t) &= \sum_{\alpha=1}^M \blambda_\alpha \braket<\Psi(t)| \omega_{\alpha} p _{\alpha} \hat{n}(\br) |\Psi(t)> ,
\end{align*}
are the forces the photons exert on the electron density \cite{tokatly2013}.
	Here $\rho_2(\br,\br',t) = \tfrac{1}{2} \sum_{k \ne l} \braket<\Psi(t)|\delta(\br-\br_k)\delta(\br'-\br_l)|\Psi(t)>$ is the pair density.
	In contrast the dressed \ac{KS} equation of motion (also for approximate $v'_\mathrm{Mxc}$) reads,
\begin{align*}
\tfrac{\partial^2}{\partial t^2} n'(\bz,t) &=
\nabla_\bz \cdot \left\{ n'(\bz,t) \nabla_\bz [v'(\bz,t) + v'_\mathrm{Mxc}(\bz,t)] \right\} \\
&- \nabla_\bz \cdot \mathbf{Q}'_\mathrm{KS}(\bz,t) ,
\end{align*}
where $\mathbf{Q}'_\mathrm{KS}(\bz,t) = \imagi \braket<\Phi'(t)|[\hat{T}',\hat{\bj}'(\bz)]|\Phi'(t)>$ and $\hat{\bj}'(\bz) = \tfrac{1}{2\imagi} \sum_{k=1}^N [\delta(\bz-\bz_k) \overrightarrow\nabla_{\bz_k} - \overleftarrow\nabla_{\bz_k} \delta(\bz-\bz_k)]$.
	Inserting the expression for $v'(\bz,t)$ and integrating this equation over all $q$-coordinates, we find (again also for approximate $v'_\mathrm{Mxc}$)
\begin{align*}
\!\!\! \tfrac{\partial^2}{\partial t^2} n(\br,t) \!&=\!
\nabla_\br \!\cdot\! \left[ n(\br,t) \nabla_\br v(\br,t) \!+\!\!\! \int \!\! \diff^M q \, n'(\bz,t) \nabla_\br v'_\mathrm{Mxc}(\bz,t) \!\right] \\
&-\nabla_\br \cdot \left[ \bF_\mathrm{dip,KS}^\mathrm{d}(\br,t) + \bF_\mathrm{lin,KS}^\mathrm{d}(\br,t) + \mathbf{Q}_\mathrm{KS}^\mathrm{d}(\br,t) \right] ,
\end{align*}
where $\bF_\mathrm{dip,KS}^\mathrm{d}(\br,t) = - \sum_{\alpha=1}^M \blambda_\alpha n(\br t) (\blambda_\alpha \cdot \br)$, $\bF_\mathrm{lin,KS}^\mathrm{d}(\br,t)
= \sum_{\alpha=1}^M \blambda_\alpha \int \diff^M q \tfrac{\omega_\alpha q_\alpha}{\sqrt{N}} n'(\bz,t) = \tfrac{1}{N} \bF_\mathrm{lin}(\br,t)$ and $\mathbf{Q}_\mathrm{KS}^\mathrm{d}(\br,t) = \imagi \braket<\Phi'(t)|[\hat{T},\hat{\bj}(\br)]|\Phi'(t)>$.
	That is, we get the $v(\br,t)$ term as well as parts of $\bF_\mathrm{dip}(\br,t)$ and $\bF_\mathrm{lin}(\br,t)$ already from $v'(\bz,t)$, even if we set $v'_\mathrm{Mxc}(\bz,t)=0$.
	We also get the kinetic forces $\mathbf{Q}_\mathrm{KS}^\mathrm{d}(\br,t)$, and $v'_\mathrm{Mxc}(\bz,t)$ then has to provide the rest of all the forces.
	To reproduce the exact $n'(\bz,t)$, $v'_\mathrm{Mxc}(\bz,t)$ should reproduce the exact forces for $n'(\bz,t)$.
	However, given our interest in physical observables, it is usually sufficient if we can get the right forces for $n(\br,t)$, which different $v'_\mathrm{Mxc}(\bz,t)$ can achieve.
	To model the forces from $\hat{W}'$ that we are clearly missing, we can approximate them by their \ac{KS} values \cite{ruggenthaler2009} using
\begin{align*}
\nabla_\bz \cdot \left[ n'(\bz,t) \nabla_\bz v'_\mathrm{Mx}(\bz,t) \right] = - \nabla_\bz \cdot \imagi \braket<\Phi'(t)|[\hat{W}',\hat{\bj}'(\bz)]|\Phi'(t)> ,
\end{align*}
which we call \ac{Mx}.
	In general, we expect that designing $v'_\mathrm{Mxc}(\bz,t)$ to approximate $\bF_\mathrm{dip}(\br,t)$ and $\mathbf{Q}(\br,t)$ works the same as in standard \ac{KS}, since they only depend on the electrons.
	So the main difference is in how to approximate $\bF_\mathrm{lin}(\br,t)$, and here we specifically get $\tfrac{1}{N} \bF_\mathrm{lin}(\br,t)$ from $v'(\bz,t)$.
	We thus need only approximate the remaining part.
	Also, to do this, we may try to scale the part of $v'(\bz,t)$ responsible for this $\tfrac{1}{N}$ contribution, $-\!\sum_{\alpha=1\!}^M \!\tfrac{\omega_\alpha q_\alpha}{\sqrt{N}}(\blambda_\alpha \!\cdot \br)$, by $N$ to get the exact force expression for $\bF_\mathrm{lin}(\br,t)$.
	Keep in mind that this not necessarily leads to the exact forces though since this would demand the exact $n'(\bz,t)$, and this idea is only designed with $n(\br,t)$ not $n'(\bz,t)$ in mind.
	Since $\bF_\mathrm{lin}(\br,t)$ depends strongly on the electron-photon correlations in $n'(\bz,t)$ it can be very sensitive to this.
	We investigate this approximation further in Supplemental Material Sec.~V, while we focus in the following on the \ac{Mx} approximation.
	Also note that the standard version of \ac{KS} translates to a special case of our dressed version (see Supplemental Material Sec.~IV).%
\footnote{%
	This follows the exact same recipe adding $\hat{H}_\mathrm{aux}$ to $\hat{H}_\mathrm{KS}(t)$ and using $\Phi'(t) = \Phi(t)\chi(t)$, with $\hat{H}_\mathrm{KS}(t)$ and $\Phi(t)$ the standard \ac{KS} Hamiltonian and wave function.
	It is somewhat different though, as it reproduces the $n'(\bz,t)$ of $\Phi'(t) = \Phi(t)\chi(t)$ instead of $\Psi'(t) = \Psi(t)\chi(t)$, but this $n'(\bz,t)$ also yields the same $n(\br,t)$ and $p_\alpha(t)$.
}
	This allows us to rewrite all standard approximations as dressed, but not vice versa.


	In the following we compare dressed \ac{Mx} with available standard functionals for two examples both featuring two particles in a singlet state coupled to one mode with no external current $J(t) = 0$.
	Therefore we have $v'(\br,q,t) = v(\br,t) + \tfrac{\omega^2}{2} q^2 - \tfrac{\omega}{\sqrt{2}} q (\blambda \cdot \br) + \tfrac{1}{2} (\blambda \cdot \br)^2$, and if we further take the spatial part of the dressed \ac{KS} wave function to be described by a doubly-occupied dressed orbital $\varphi'(\br,q,t)$, we have explicitly that
\begin{align*}
v'_\mathrm{Mx}(\br,q,t) &= \tfrac{1}{2} \int \diff^3 \br' \, n(\br',t) w(\br,\br') \\
&+ \tfrac{1}{2} \left\{ [\blambda \cdot \mathbf{R}(t) - \omega p(t)](\blambda \cdot \br) - \tfrac{\omega}{\sqrt{2}} q \blambda \cdot \mathbf{R}(t) \right\} .
\end{align*}
	Here $\mathbf{R}(t) \!=\! \int \diff^3 r \, \br n(\br,t)$ is the total dipole, and we recall $n(\br,t)$ and $p(t)$ are given in terms of $n'(\bz,t) = 2|\varphi'(\bz,t)|^2$.
	Interestingly, the second line equals half the standard \ac{KS} mean-field contribution in this particular case (while $v'(\bz,t)$ gives a different approximation for the other half).
	The first line is the Hartree exact exchange potential $v_\mathrm{Hx}(\br,t)$, which here equals half the Hartree potential.
	In symmetric cases (like our second example), also for more electrons and modes, $v'_\mathrm{Mx}(\bz,t) = v_\mathrm{Hx}(\br,t)$, and the coupling to the photon mode is only due to $v'(\bz,t)$.

\begin{figure} [H]
\includegraphics[width=8.6cm]{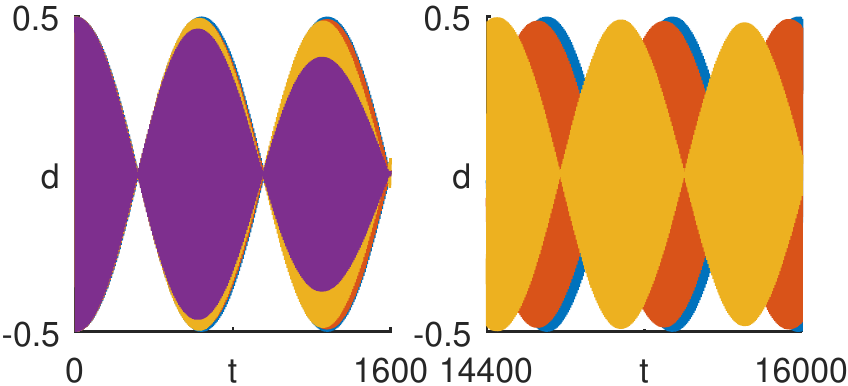}
\caption{(color online) The exact (orange), \acs*{TDOEP}-($GW_0$) (lilac), standard \ac{Mx} (blue) and dressed \ac{Mx} (red) dipole moment of two electrons on two sites coupled to one mode.}
\label{fig:TwoSite}
\end{figure}

	In our first example, we consider a two-site model with $v = w = 0$,
\begin{align*}
\hat{H}\! = \! - t \sum_{\sigma=1}^2 (\hat{c}_{1,\sigma}^\dagger \hat{c}_{2,\sigma} + \hat{c}_{2,\sigma}^\dagger \hat{c}_{1,\sigma}) + \tfrac{1}{2} \left[ -\tfrac{\partial^2}{\partial p^2} + \left(\omega p \!-\! \lambda \hat{d} \right)^2 \right] .
\end{align*}
	Here $\hat{c}_{k, \sigma}^\dagger$ and $\hat{c}_{k, \sigma}$ are fermionic creation and annihilation operators of the electrons at site $k$ with spin $\sigma$, while $\hat{d} = \tfrac{1}{2} \sum_{\sigma=1}^2 (\hat{c}_{2, \sigma}^\dagger \hat{c}_{2, \sigma} - \hat{c}_{1, \sigma}^\dagger \hat{c}_{1, \sigma})$ is the dipole operator.
	As initial state we pick a spin-singlet electronic state and the photon mode in its uncoupled ground state, i.e., $\Psi_0 = \tfrac{1}{\sqrt{2}} \left(\sqrt{\tfrac{1}{4}} \; \hat{c}_{1,\uparrow}^\dagger + \sqrt{\tfrac{3}{4}} \; \hat{c}_{2,\uparrow}^\dagger \right) \left( \sqrt{\tfrac{1}{4}} \; \hat{c}_{1,\downarrow}^\dagger + \sqrt{\tfrac{3}{4}} \; \hat{c}_{2,\downarrow}^\dagger \right) \ket|0>_\mathrm{e} \! \otimes \! \ket|0>_\mathrm{p}$.
	Here $\ket|0>_\mathrm{e}$ and $\ket|0>_\mathrm{p}$ refer to the electronic and photonic vacuum (the ground state of the harmonic oscillator in displacement representation).
	The resulting exact Rabi oscillation for $t=0.5$, $\omega=1$ and $\lambda = 0.01$ is depicted in \cref{fig:TwoSite} in orange.
	We compare this to dressed \ac{Mx} using $\Phi'_0 = \Psi_0 \otimes \ket|0>_{ p_2}$ as initial state, where $\ket|0>_{ p_2}$ is the ground state of $\hat{H}_\mathrm{aux}$, and to three cases within the standard \ac{KS} scheme using $\Psi_0$ as initial state.
	Dressed \ac{Mx} improves upon its direct counterpart standard \ac{Mx} (blue) and standard mean-field (not plotted as it is visually indistinguishable from \ac{Mx} for the given times), which especially manifests for later times.
	The third standard scheme is a \ac{TDOEP} $GW_0$ approach based on the exact-exchange energy expression of the Lamb shift \cite{pellegrini2015}.
	This relatively advanced functional accurately describes the correlated ground state \cite{pellegrini2015,flick2018ab}, but does not self-consistently incorporate the polaritonic eigenstates into the non-adiabatic potential, which manifests as a beating (lilac).
	In contrast the non-perturbative character of dressed \ac{Mx} also allows the application to the ultra-strong coupling regime (e.g., circuit quantum electrodynamics), still with a consistent improvement over standard \ac{Mx} and mean-field for all $\lambda$ we tested.

\begin{figure} [H]
\includegraphics[width=8.6cm]{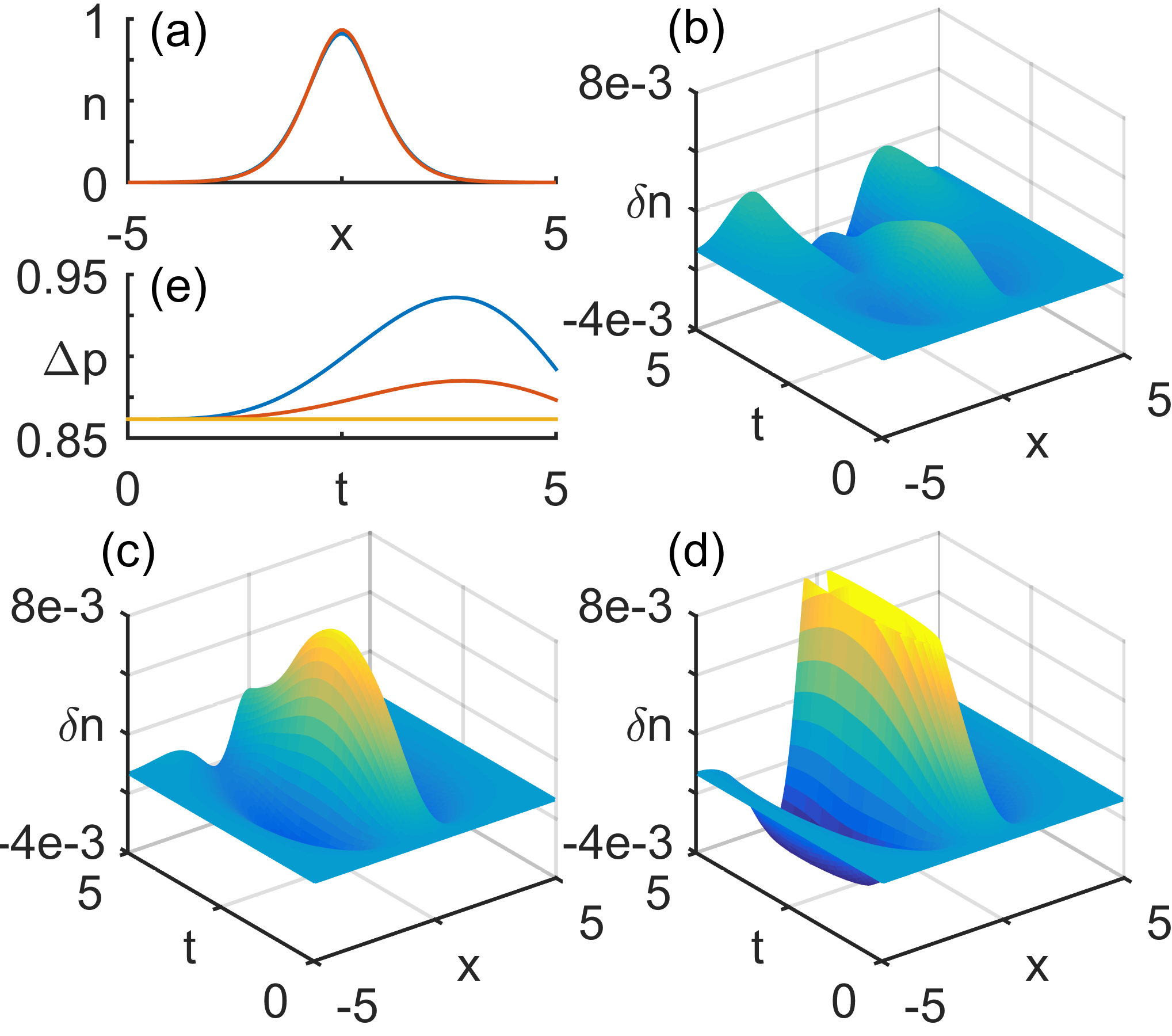}
\caption{(color online) (a) The exact (blue) and \ac{Mx} (red) ground-state densities of the bare ($\lambda=0$) one-dimensional Helium model. The corresponding changes in the (b) exact (c) dressed \ac{Mx} and (d) standard \ac{Mx} density, $\delta n(x,t) = n(x,t) - n_0(x)$, when placed inside a cavity, and (e) the exact (blue), dressed \ac{Mx} (red) and standard \ac{Mx} (orange) field fluctuations.}
\label{fig:Helium}
\end{figure}

	In our second example we go beyond a simple two-site model, and consider a one-dimensional model of helium using the soft Coulomb interaction, $v(x) = -2/\sqrt{x^2+1}$ and $w(x,x') = 1/\sqrt{|x-x'|^2+1}$ \cite{ruggenthaler2009b, fuks2011}, and with hard wall boundary conditions at $x = \pm 5$.
	Here we want to investigate true quantum-induced light-matter dynamics in real space where classical mean-field methods will fail \cite{hoffmann2019capturing}.
	We first determine the exact ground-state $\psi_0$ of the ``bare'' helium (see \cref{fig:Helium} (a)), i.e, outside of the cavity, which is an electronic spin-singlet state.
	The excitation frequency for the lowest excited state is $\omega_1 = 0.58037$.
	Upon bringing helium into an empty cavity with $\omega = \omega_1$ and $\lambda=0.1$, i.e., we use a tensor product $\Psi_0 = \psi_0 \otimes \ket|0>_\mathrm{p}$, we find the spontaneous emission behaviour in \cref{fig:Helium} (b).
	As we only have one mode the emitted energy cannot dissipate and reoccurance takes place \cite{flick2017a}.
	To investigate how the dressed \ac{KS} scheme performs we first solve for the bare \ac{KS} ground state $\varphi_0$ using the \ac{Mx} approximation for $\lambda=0$ (see \cref{fig:Helium} (a)) and then set up the corresponding matter-photon initial state $\Phi'_0 = \varphi_0 \otimes \ket|0>_\mathrm{p} \otimes \ket|0>_{\mathrm{p}_2}$.
	The photon field is then chosen with the same coupling strength $\lambda$ and frequency $\omega$ as in the exact reference calculation above.
	In \cref{fig:Helium} (c) we can then see that a self-consistent propagation using the dressed \ac{Mx} approximation qualitatively recovers the photon-induced dynamics and improves upon standard \ac{Mx} in \cref{fig:Helium} (d).
	In contrast standard mean-field yields no density change as the electric field is always zero as the dipole moment is.
	Such a real-time and real-space test of an approximation is a much harder test than say, only looking at reduced quantities such as the energy or the dipole moment \cite{schafer2018ab}.
	Note that within the standard \ac{KS} scheme we solved the Coulomb interaction on an exact exchange level, and used $\Phi_0 = \varphi_0 \otimes \ket|0>_\mathrm{p}$ as initial state.
	The difference between \ac{Mx} and mean-field thus refers to the light-matter interaction.

	By construction both the exact and the dressed \ac{Mx} approximation generate the exact $p(t) = 0$ in the given example.
	However, let us finally illustrate how we can also investigate observables such as the field fluctuations $\Delta p(t) = \braket<\Psi(t)|p^2|\Psi(t)> - p^2(t)$ of the photon field that are very challenging in standard \ac{KS} \cite{flick2017a}.
	The dressed approximation to the field fluctuations generally reads $\int \diff^{3+M} z \, \tfrac{q_\alpha^2}{N} n'(\bz,t) + \int \diff^{3+M} z \, \diff^{3+M} z' \tfrac{2 q_\alpha q'_\alpha}{N} \rho'_2(\bz_1,\bz_2,t) - p_\alpha^2(t)$, which in the given case reduces to $\int \diff^2 z \, \tfrac{q^2}{2} n'(\bz,t)$.
	We find again a qualitative agreement (see \cref{fig:Helium} (e)).
	In contrast standard \ac{Mx} yields constant field fluctuations (as typical in standard \ac{KS} as changes in the fluctuations originate from explicit light-matter correlation).


	To conclude, we have presented a novel approach to coupled light-matter systems based on a higher-dimensional auxiliary system that explicitly correlates light and matter, i.e., that considers polaritons as the underlying particles.
	We exemplified the possibilities by presenting an alternative \ac{KS} construction and providing first benchmarks.
	We observed a consistent improvement of dressed \ac{Mx} over its direct counterpart (standard \ac{Mx}) as $\tfrac{1}{N}$ of $\bF_\mathrm{lin}(\br,t)$ is covered directly by the known $v'(\bz,t)$ (for exact $n'(\bz,t)$) and only the remaining part is approximated.
	Scaling this $\tfrac{1}{N}$ fraction could lead to further improvements (see Supplemental Material Sec.~V).
	Due to its non-perturbative character, our method remains applicable even for ultra-strong couplings, rendering this an ideal approach to low dimensional highly correlated problems such as circuit QED.
	Note that any reasonably generic \ac{KS} code can be used to perform the dressed \ac{KS} computations by working in higher dimensions $\bz$ and by selecting the proper orbitals (see Supplemental Material Sec.~II).
	Finally, we point out that the dressed approach can equally-well be applied beyond the context of coupled light-matter systems.
	For instance, instead of treating photon modes strongly correlated with matter one could consider strong coupling to phonon modes (polarons).
	One can even think about modelling electron-electron correlation with the help of auxiliary degrees of freedom, e.g., dispersion like interactions that are dipole-dipole interactions such as the van-der-Waals interaction.
	Such future objectives make this approach interesting also beyond the current physical context.


\textit{Acknowledgement.}

	We acknowledge financial support from the European Research Council (ERC-2015-AdG-694097) and Grupos Consolidados (IT578-13).
	We would like to thank F.\ Buchholz and J.\ Flick for useful discussions.


\bibliographystyle{apsrev4-1}
\bibliography{Library}


\end{document}


\newcommand \diff {\mathrm{d}}
\newcommand \imagi {\mathrm{i}}
\newcommand \br {\mathbf{r}}
\newcommand \bz {\mathbf{z}}
\newcommand \blambda {\bm{\lambda}}
\newcommand \bj {\mathbf{j}}
\newcommand \bF {\mathbf{F}}

\NewDocumentCommand \bra{r<|}{\langle #1 \vert}
\NewDocumentCommand \ket{r|>}{\vert #1 \rangle}
\NewDocumentCommand \braket{r<>}{\langle \StrSubstitute{#1}{|}{\vert } \rangle}
\RenewDocumentCommand \outer{r|>r<|}{\vert #1 \rangle \langle #2 \vert}
\NewDocumentCommand \Bra{r<|}{\left\langle #1 \right\vert}
\NewDocumentCommand \Ket{r|>}{\left\vert #1 \right\rangle}
\NewDocumentCommand \Braket{r<>}{\left\langle \StrSubstitute{#1}{|}{\middle\vert } \right\rangle}
\NewDocumentCommand \Outer{r|>r<|}{\left\vert #1 \middle\rangle \middle\langle #2 \right\vert}

\acuse{KS}
\acuse{Mxc}
\acuse{Mx}


\title{Supplemental:\\ Dressed-Orbital Approach to Cavity Quantum Electrodynamics and Beyond}

\author{Soeren Ersbak Bang Nielsen}
\author{Christian Sch\"afer}
\author{Michael Ruggenthaler}
\affiliation{Max Planck Institute for the Structure and Dynamics of Matter and Center for Free-Electron Laser Science, Luruper Chaussee 149, 22761 Hamburg, Germany}
\author{Angel Rubio}
\affiliation{Max Planck Institute for the Structure and Dynamics of Matter and Center for Free-Electron Laser Science, Luruper Chaussee 149, 22761 Hamburg, Germany}
\affiliation{Center for Computational Quantum Physics (CCQ), The Flatiron Institute, 162 Fifth avenue, New York NY 10010}

\date{\today}

\maketitle

\section{Auxiliary Coordinate Transformation} \label{sec:Transformations}

	Here we discuss the class of transformations allowing us to rewrite the physical problem in a higher-dimensional configuration space.
	We first specify this class abstractly, as many different transformations are valid, as long as they satisfy a few basic properties.
	Although the specific choice of transformation does not matter we also present one possible explicit option for completeness and understanding.
	Finally, we also show how the physical density $n(\br,t)$ and photon-displacement $p_\alpha(t)$ are given in terms of the dressed density $n'(\bz,t)$.

\subsection{Abstract Transformation}

	To establish a transformation from $p_k$ to $q_k$ coordinates that is orthogonal and also satisfies $p=\tfrac{1}{\sqrt{N}}(q_1+...+q_N)$, all we need to do is to require that the $p$-axis lies along the unit vector $\tfrac{1}{\sqrt{N}}(1,...,1)$ in $q$-space, while the other axes must simply be orthogonal to this and each other.
	If we further keep the same scale for all $p_k$ and $q_k$ axes, we have our desired transformation.

\subsection{Explicit Transformation}

	To give one explicit special case of the above class of transformations, we may use the following for $4$ electrons,
\begin{align*}
p     &= \tfrac{1}{\sqrt{ 4}} \left( q_1 + q_2 + q_3 + q_4 \right) , \\
p_2 &= \tfrac{1}{\sqrt{ 2}} \left( q_1 - q_2 \right) , \\
p_3 &= \tfrac{1}{\sqrt{ 6}} \left( q_1 + q_2 - 2q_3 \right) , \\
p_4 &= \tfrac{1}{\sqrt{12}} \left( q_1 + q_2 + q_3 - 3q_4 \right) ,
\end{align*}
which clearly has an orthogonal transformation matrix.
	This transformation is easily generalised to any number of electrons using
\begin{align*}
p_k = \tfrac{1}{\sqrt{k^2-k}} \left( q_1 + ... + q_{k-1} - [k-1] q_k \right)
\end{align*}
for $2 \le k \le N$, of course, alongside $p = \tfrac{1}{\sqrt{N}} (q_1+...+q_N)$.
	The inverse of an orthogonal matrix is just the transpose, so for example for 4 electrons the inverse transformation (the one actually used to go from $p$- to $q$-coordinates) is
\begin{align*}
q_1 &= \tfrac{1}{\sqrt{4}} p + \tfrac{1}{\sqrt{2}} p_2 + \tfrac{1}{\sqrt{6}} p_3 + \tfrac{1}{\sqrt{12}} p_4 \\
q_2 &= \tfrac{1}{\sqrt{4}} p - \tfrac{1}{\sqrt{2}} p_2 + \tfrac{1}{\sqrt{6}} p_3 + \tfrac{1}{\sqrt{12}} p_4 \\
q_3 &= \tfrac{1}{\sqrt{4}} p - \tfrac{2}{\sqrt{6}} p_3 + \tfrac{1}{\sqrt{12}} p_4 \\
q_4 &= \tfrac{1}{\sqrt{4}} p - \tfrac{3}{\sqrt{12}} p_4
\end{align*}

\subsection{Wave Function Relations}

	For normalised $\Psi(\br_1 \sigma_1, ... , \br_N \sigma_N, p_1, ..., p_M, t)$ and $\chi(p_{1,2}, ..., p_{M, N},t)$ and $\Psi'(\bz_1 \sigma_1, ..., \bz_N \sigma_N, t) \equiv \Psi\chi$ we find that
\begin{align*}
1 &= \sum_{\sigma_1,...,\sigma_N} \int \diff^{3N} r\, \diff^{MN} p \, |\Psi \chi|^2 \\ 
&=  \sum_{\sigma_1,...,\sigma_N} \int \diff^{(3+M)N} z\, |\Psi'|^2 ,
\end{align*}
so the total wave function $\Psi'$ is also normalised.
	For $\Psi'$ properly anti-symmetrized we further find that,
\begin{align}
n(\br,t) &= N \sum_{\sigma_1,...,\sigma_N} \int \diff^{3(N-1)} r \, \diff^{MN} p \, |\Psi \chi|^2 \nonumber \\
&= \int \diff^M q \underbrace{N \sum_{\sigma_1,...,\sigma_N} \int \diff^{(3+M)(N-1)} z \, |\Psi'|^2}_{= n'(\bz,t)} , \nonumber \\
\label{pAlpha}
p_\alpha(t) & = \sum_{\sigma_1,...,\sigma_N} \int \diff^{3N} r \, \diff^{MN} p \, p_\alpha |\Psi \chi|^2 \\
&= \int \diff^{3+M} z \, \tfrac{q_\alpha}{\sqrt{N}} n'(\bz,t) . \nonumber
\end{align}
	For later convenience, let us finally introduce the physical and dressed two-particle densities \cite{stefanucci2013},
\begin{align*}
\rho_2(\br,\br',t) &= \tfrac{N(N-1)}{2} \sum_{\sigma_1,...,\sigma_N} \int \diff^{3(N-2)} r \diff^{M} p \, |\Psi|^2 , \\
\rho_2'(\bz,\bz',t) &= \tfrac{N(N-1)}{2} \sum_{\sigma_1,...,\sigma_N} \int \diff^{(3+M)(N-2)} z \, |\Psi'|^2 ,
\end{align*}
which are related by $\rho_2(\br,\br',t) = \int \diff^M q \diff^M q' \rho_2'(\bz,\bz',t)$.

\section{Symmetries} \label{sec:Symmetries}

	In this \namecref{sec:Symmetries} we consider the space of all dressed Hamiltonians $\hat{H}'(t)$ characterised by $v'(\bz,t)$ and $w'(\bz,\bz')$, which is displayed in \cref{fig:DressedHamiltonianSpace}.
\begin{figure} [H]
\begin{tikzpicture}[scale=1, font=\small]
\tikzmath{\xmin=-4.3; \xmid=0.0; \xmax=4.3; \ymin=-3.7; \ymid=0.0; \ymax=1.0; \indentation=0.00;}
\draw (\xmin,\ymin) rectangle (\xmax,\ymax);
\draw[dashed] (\xmin+0.02,\ymid) -- (\xmax-0.02,\ymid)
node[pos=0.0675,fill=white,font=\tiny]{$\cup$}
node[pos=0.19,fill=white,font=\tiny]{$\cup$}	
node[pos=0.315,fill=white,font=\tiny]{$\cup$}
node[pos=0.4375,fill=white,font=\tiny]{$\cup$}
node[pos=0.5625,fill=white,font=\tiny]{$\cup$}
node[pos=0.685,fill=white,font=\tiny]{$\cup$}
node[pos=0.81,fill=white,font=\tiny]{$\cup$}
node[pos=0.9325,fill=white,font=\tiny]{$\cup$};
\draw[dashed] (\xmid,\ymin) -- (\xmid,\ymax)
node[pos=0.1,fill=white,font=\tiny] {$\supset$}
node[pos=0.3,fill=white,font=\tiny] {$\supset$}
node[pos=0.5,fill=white,font=\tiny] {$\supset$}
node[pos=0.7,fill=white,font=\tiny] {$\supset$}
node[pos=0.9,fill=white,font=\tiny] {$\supset$};
\node[below right, align=left] at (\xmin+\indentation,\ymax-\indentation)
{All $\hat{H}'[v',w']$\\
$\circ$ $\bz_k \sigma_k \leftrightarrow \bz_l \sigma_l$ symmetry};
\node[below right, align=left] at (\xmin+\indentation,\ymid-\indentation)
{Physical $\hat{H}'[v,w,\omega_\alpha,\blambda_\alpha,\dot{J}_\alpha]$ \\
$\rightarrow$ The physical properties: \\
$\bullet$ $\hat{H} + \sum_\alpha \hat{H}_{\mathrm{aux},\alpha}$ \\
$\bullet$ $\Psi'=\Psi \prod_\alpha \chi_\alpha$ \\
$\circ$ $\br_k \sigma_k \leftrightarrow \br_l \sigma_l$ symmetry \\
$\circ$  $q_{\alpha,k} \leftrightarrow q_{\alpha,l}$ symmetry};
\node[below right, align=left] at (\xmid+\indentation,\ymid-\indentation)
{Physical $\hat{H}'[v,\omega_\alpha,\dot{J}_\alpha]$ with \\
$w'=0$, i.e., $w=\blambda_\alpha=0$. \\
$\rightarrow$ \parbox{5cm}{The non-interacting \\ physical properties:} \\
$\bullet$ $\hat{H}_\mathrm{E} + \sum_\alpha (\hat{H}_{p_\alpha} + \hat{H}_{\mathrm{aux},\alpha})$ \\
\hspace{2.4mm} - $v'(\bz) = v(\br) \!+\! \sum_\alpha \!v_{q_\alpha}(q_\alpha)$ \\
$\bullet$ $\Psi' = \Psi_\mathrm{E} \prod_\alpha \varphi_{p_\alpha} \chi_\alpha$ \\
\hspace{2.4mm} - $n'(\bz) = n(\br) \prod_\alpha n_{q_\alpha}(q_\alpha)$ \\
\textendash~Includes the standard $\hat{H}'_\mathrm{KS}$ };
\node[below right, align=left] at (\xmid+\indentation,\ymax-\indentation)
{All $\hat{H}'[v']$ with $w'=0$ \\
\textendash~Includes the dressed $\hat{H}'_\mathrm{KS}$ };
\end{tikzpicture}
\caption{The space of all dressed Hamiltonians $\hat{H}'$.\\
$\bullet$ Separation of $\hat{H}'$ and the corresponding eigenstates $\Psi'$.\\
$\circ$ Exchange symmetries (silently also apply for all subsets $\subset$).\\
Note that $\hat{H}, \hat{H}_\mathrm{E}, \hat{H}_{p_\alpha}, \hat{H}_{\mathrm{aux},\alpha}, \Psi, \Psi_\mathrm{E}, \varphi_{p_\alpha}$ and $\chi_\alpha$ each satisfy all exchange symmetries (with $\bz_k\sigma_k \leftrightarrow \bz_l\sigma_l$ and $\br_k\sigma_k \leftrightarrow \br_l\sigma_l$ anti-symmetry for $\Psi$ and $\Psi_\mathrm{E}$ but symmetry in all other cases).}
\label{fig:DressedHamiltonianSpace}
\end{figure}
	Only a subset of these $\hat{H}'(t)$ corresponds to a physical Hamiltonian, $\hat{H}'(t) = \hat{H}(t) + \sum_{\alpha=1}^M \hat{H}_{\mathrm{aux},\alpha}$, characterised by $v(\br,t)$, $w(\br,\br')$, $\omega_\alpha$, $\blambda_\alpha$ and $\dot{J}_\alpha(t)$ through
\begin{align}
\label{DressedPhysicalPotential}
v'(\bz,t) &= v(\br,t) + \sum_{\alpha=1}^M \Big[
\tfrac{1}{2} \omega_\alpha^2 q_\alpha^2 - \tfrac{\omega_\alpha}{\sqrt{N}} q_\alpha (\blambda_\alpha \cdot \br) \\
&+ \tfrac{1}{2} (\blambda_\alpha \cdot \br)^2 + \tfrac{\dot{J}_\alpha(t) q_\alpha}{\sqrt{N} \omega_\alpha} \Big] , \nonumber \\
\label{DressedPhysicalInteraction}
w'(\bz,\bz') &= w(\br,\br') + \sum_{\alpha=1}^M \Big[
- \tfrac{\omega_\alpha}{\sqrt{N}} q_\alpha (\blambda_\alpha \cdot \br') \\
&- \tfrac{\omega_\alpha}{\sqrt{N}} q'_\alpha (\blambda_\alpha \cdot \br)
+ (\blambda_\alpha \cdot \br)(\blambda_\alpha \cdot \br') \Big] , \nonumber 
\end{align}
while all other pairs of $v'(\bz,t)$ and $w'(\bz,\bz')$ correspond to a non-physical $\hat{H}'(t)$, and form an extension of the space.
	In \cref{sec:SymmetriesOfDressedPhysicalSystems} we show that this subset of $\hat{H}'(t)$ exhibits certain properties that most non-physical $\hat{H}'(t)$ do not, and so we call these ``the physical properties'' in \cref{fig:DressedHamiltonianSpace}.
	That is, these properties rely on a fine balance between $v'(\bz,t)$ and $w'(\bz,\bz')$ that approximations can break, e.g., it is important $v'(\bz,t)$ and $w'(\bz,\bz')$ use the same $\omega_\alpha$ and $\blambda_\alpha$ or $\hat{H}'(t)$ is not physical, and will break the properties.
	We further show that for a time-independent physical $\hat{H}'$ the eigenfunctions separate in $\Psi' = \Psi \prod_{\alpha=1}^M \chi_\alpha$, where $\Psi$ and each $\chi_\alpha$ satisfy all the physical exchange symmetries.
	There are also many non-physical eigenstates not on this form, so to compute the physical eigenstates directly from $\hat{H}'$, one must enforce the physical properties, or compute all eigenstates and extract those on this form afterwards.
	The subset of $\Psi'_0$ corresponding to a physical $\Psi_0$ also satisfies the physical separability and exchange symmetries.
	In this way, the dressed system is by design equivalent with the physical system when $v'(\bz,t)$, $w'(\bz,\bz')$ and $\Psi'_0$ correspond to a physical system, but at the same time it is also an extension to any $v'(\bz,t)$, $w'(\bz,\bz')$ and $\Psi'_0$.
	To end \cref{sec:SymmetriesOfDressedPhysicalSystems} we also discuss the all $\blambda_\alpha=0$ case, where $\hat{H}'(t)$ separates in an electron and photon parts so \ac{KS} should usually also reduce to purely electronic \ac{KS}.
	
	The dressed \ac{KS} scheme is based on the extended space, and comprises a $\Phi'_0$ and $v'_\mathrm{KS}(\bz,t)$ that recreate the $n'(\bz,t)$ of the physical $\Psi'(t)$, and thereby $n(\br,t)$ as well as $p_\alpha(t)$.
	In \cref{sec:SymmetriesOfKohnShamSystems} we will thus show that any dressed $\Phi'(t)$ has none of the physical properties (unless all $\blambda_\alpha = 0$), and therefore does not correspond to any physical $\Phi'(t)$.
	This difference of $\Phi'(t)$ and the $\Psi'(t)$ that it recreates the density of is compensated by $v'_\mathrm{Mxc}(\bz,t)$.
	It has to be build into approximate $v'_\mathrm{Mxc}(\bz,t)$ to do the same though.
	More generally we may consider any $\Phi'_0$ and $v'_\mathrm{KS}(\bz,t)$ that recreate $n(\br,t)$ and $p_\alpha(t)$, but different $n'(\bz,t)$.
	These are all equally interesting to the extent we can obtain them.%
\footnote{%
	Essentially, the only minor difference is how well $\Phi'(t)$ directly reproduce other quantities not given in terms of $n(\br,t)$ and $p_\alpha(t)$.
}
	An important special case is the standard version \cite{tokatly2013} of cavity \ac{QEKS}, with \ac{KS} wave function $\Phi(t)$.
	There is really no practical reason to perform standard \ac{KS} in the dressed setting, yet it is worth our study to gain insight into dressed \ac{KS} and for comparison.
	In \cref{sec:RelationshipWithStandardKS} we thus show that standard \ac{KS} corresponds to recreate the $n'_{\!}(\bz,_{\!}t)$ of $\Phi'(t) \!= \Phi(t) \chi(t)$, and the standard \ac{KS} Hamiltonian $\hat{H}'_\mathrm{KS}(t)$ takes the same form as a non-interacting ($w(\br,\br')=\blambda_\alpha=0=w'(\bz,\bz')$) physical $\hat{H}'(t)$ defined by $v(\br,t)$, $\omega_\alpha$ and $\dot{J}_\alpha(t)$.	
	That is, the standard \ac{KS} potential can be written as \scalebox{0.96}[1]{$v'_\mathrm{KS}(\bz,t) \!=\! v_{\mathrm{KS},\br}(\br,t) + \sum_{\alpha=1}^M [\tfrac{1}{2} \omega_\alpha^2 q_\alpha^2 +\! \tfrac{\dot{J}_{\mathrm{KS},\alpha}(t) q_\alpha}{\sqrt{N} \omega_\alpha}]$} to match the physical $v'(\bz,t) = v(\br,t) + \sum_{\alpha=1}^M [\tfrac{1}{2} \omega_\alpha^2 q_\alpha^2 + \tfrac{ \dot{J}_\alpha(t) q_\alpha}{\sqrt{N} \omega_\alpha}]$.
	This form of the potentials ensures the further properties for non-interacting physical systems presented in \cref{fig:DressedHamiltonianSpace}, i.e., further separations in an electron and photon parts.
	That is, we see that $v'_\mathrm{KS}(\bz,t)$ separates in an electron and photon parts and therefore so does $\hat{H}'_\mathrm{KS}(t)$ and so $\hat{H}_\mathrm{KS}(t)$.
	Thus \scalebox{0.94}[1]{$\hat{H}'_\mathrm{KS}(t) \!=\! \hat{H}_\mathrm{KS,E}(t) + \sum_{\alpha=1}^M [\hat{H}_{\mathrm{KS},p_\alpha}(t) + \hat{H}_{\mathrm{aux},\alpha}]$}.
	\!Also, for a time-independent $\hat{H}'_\mathrm{KS}$ the eigenstates thus take the form $\Phi' = \Phi_\mathrm{E} \prod_{\alpha=1}^M \varphi_{p_\alpha} \chi_\alpha$ for which the corresponding dressed densities separate $n'(\bz) = n(\br) \prod_{\alpha=1}^M n_{q_\alpha}(q_\alpha)$.%
\footnote{%
	Note that $n_{q_\alpha}(q_\alpha) = \int \diff^{3N} r \, \diff^{NM-1} q |\Psi'|^2$ are not the photon mode densities $n_{p_\alpha}(p_\alpha) = \int \diff^{3N} r \, \diff^{M-1} p |\Psi|^2$.
}
	This form tells us how to construct the $\Phi'$ eigenstates if we first compute the $\br\sigma$-orbitals of $\Phi_\mathrm{E}$, the $p_\alpha$-orbitals and the $p_{\alpha,k}$-orbitals of $\chi_\alpha$ using (the one-body form of) $\hat{H}_\mathrm{KS,E}$, $\hat{H}_{\mathrm{KS},p_\alpha}$ respectively $\hat{H}_{\mathrm{aux},\alpha}$.
	The exception is that it is beyond our scope to determine the excited $\chi_\alpha$, which depend on $p_{\alpha,2},...,p_{\alpha,N}$ yet have $q_{\alpha,k} \leftrightarrow q_{\alpha,l}$ symmetry, as we are primarily interested in the ground state anyway.
	In \cref{sec:SymmetriesOfKohnShamSystems} we show that the $\Phi'$ can also be written as fixed linear combinations of determinants of $\bz\sigma$-orbitals (already in \cref{sec:SymmetriesOfDressedPhysicalSystems} we study how $\varphi_{p_\alpha} \chi_\alpha$ look in $q_{\alpha,k}$ coordinates), which tells us how to construct the $\Phi'$ if we instead first compute the $\bz\sigma$-orbitals using the full $\hat{H}'_\mathrm{KS}$.
	We do not determine all the specific linear combinations, but we do show how it is straightforward in simple cases (like lowly excited $\Phi_\mathrm{E}$ and $\varphi_{p_\alpha}$) as are most relevant to us.
	In fact we only need the ground state in ground state \ac{KS}, but low excitations of $\Phi_\mathrm{E}$ and $\varphi_{p_\alpha}$ may also be relevant to for example compute further properties, and are anyway worth a study to better understand the dressed structure.
	We show that the ground state is typically just a single determinant (or linear combination of a few differing only in their spin function).
	However, one has to use the right $\bz\sigma$-orbitals, which are often not those with lowest energy.
	For excited states the number of determinants grows fast with the excitation levels, but at least for few modes and low excitations the number of different orbitals remains manageable even so, allowing practical use in many cases.
	In dressed \ac{KS} we may use the same linear combinations of determinants for the $\Phi'$ as we do in standard \ac{KS}, only, as $v'_\mathrm{KS}(\bz)$ gets correlated for $\blambda_\alpha \ne 0$ so do the $\bz\sigma$-orbitals.
	For example, we show that using the same form for the \ac{KS} ground state in dressed and standard ground state \ac{KS} ensures that both theories coincide and reduce to purely electronic \ac{KS} for physical systems with all $\blambda_\alpha=0$, as we show they should as they are to recreate the same $n'(\bz)$.
	If further $w(\br,\br')=0$ for the physical system, this also ensures that the physical and both \ac{KS} systems coincide.
	Using the same forms also for the excited states likewise ensures equivalence.
	These are strong arguments to use the standard \ac{KS} forms for the $\Phi'$ also in dressed \ac{KS} as other forms would break these equivalences and mimic the states by different forms instead of by the right ones.
	For those $\Phi'$ that consist of multiple determinants, the degeneracy of the determinants is often lifted in dressed \ac{KS} though due to the correlated orbitals (an important exception is if the determinants only differ in their spin).
	These $\Phi'$ are therefore no longer truly eigenstates of $\hat{H}'_\mathrm{KS}$ in dressed \ac{KS} (unless all $\blambda_\alpha=0$), so this is a price of we \emph{want} the equivalences as clarified further in \cref{sec:SymmetriesOfKohnShamSystems}.
	It is therefore not without issues to break the symmetries as dressed \ac{KS} do, but dressed ground state \ac{KS} seems to go free of these issues as long as $\Phi'$ is a single determinant (or the determinants only differ in their spin), as such $\Phi'$ are still eigenstates of $\hat{H}'_\mathrm{KS}$.
	Overall, standard \ac{KS} thus has a clear advantage in that $\Phi'(t)$ by design is physical (and we can also solve it without the dressed description).
	However, if correlated orbitals, which rely on breaking the physicality,%
\footnote{\label{foot:premise}%
	Insisting on a physical $v'_\mathrm{KS}(\bz,t)$ in the sense of \cref{DressedPhysicalPotential,DressedPhysicalInteraction} leads to standard \ac{KS} (and separating orbitals), as $v'_\mathrm{KS}(\bz,t)$ must (for a non-interacting $w'_\mathrm{KS}(\bz,\bz')=0$ \ac{KS} system) then be of the non-interacting physical form of standard \ac{KS}, for which only the standard \ac{KS} $v_{\mathrm{KS},\br}(\br,t)$ and $\dot{J}_{\mathrm{KS},\alpha}(t)$ recreate $n(\br,t)$ and $p_\alpha(t)$.
	The premise of using other $n'(\bz,t)$ and correlated orbitals is thus to break the physical properties (see also \cref{foot:correlation}, where we argue that $v'_\mathrm{KS}(\bz,t)$ must separate for $\hat{V}'_\mathrm{KS}(t)$ to have the physical $\br_k \sigma_k \leftrightarrow \br_l \sigma_l$ and $q_{\alpha,k} \leftrightarrow q_{\alpha,l}$ exchange symmetries).
}
capture the electron-photon correlation better, the corresponding $\Phi'(t)$ may well still be closer to $\Psi'(t)$ overall.
	Also, lastly we show that even if $\Phi'(t)$ is not physical, many derived quantities like the corresponding electron $n$-body densities recover appropriate properties.

\subsection{Symmetries of Dressed Physical Systems} \label{sec:SymmetriesOfDressedPhysicalSystems}

	Since electrons are fermions, $\hat{H}(t)$ is symmetric under exchange of $\br_k \sigma_k$ and $\br_l \sigma_l$.
	It is further symmetric under exchange of $q_{\alpha,k}$ and $q_{\alpha,l}$ when written in terms of $q_{\alpha,k}$ instead of the $p_\alpha$, as the $p_\alpha$ by design have this symmetry.
	Since further all $\hat{H}_{\mathrm{aux},\alpha}$ share these exchange symmetries, $\hat{H}'(t) = \hat{H}(t) + \sum_{\alpha=1}^M \hat{H}_{\mathrm{aux},\alpha}$ inherits both these as well.
	This in turn also implies that $\hat{H}'(t)$ is symmetric under exchange of $\bz_k \sigma_k$ and $\bz_l \sigma_l$, since this is just a combination of the other symmetries.
	This establishes all the physical properties of physical $\hat{H}'(t)$ presented in \cref{fig:DressedHamiltonianSpace}.

	Given some physical time-independent Hamiltonian $\hat{H}$, we may thus compute its $\br_k \sigma_k \leftrightarrow \br_l \sigma_l$ anti-symmetric eigenstates	 and the $q_{\alpha,k} \leftrightarrow q_{\alpha,l}$ symmetric eigenstates of the $\hat{H}_{\mathrm{aux},\alpha}$.
	These can then be combined into eigenstates of $\hat{H}'$ of the form $\Psi'=\Psi\prod_{\alpha=1}^M \chi_\alpha$ with both symmetries and therefore also with the dressed fermionic $\bz_k \sigma_k \leftrightarrow \bz_l \sigma_l$ anti-symmetry.
	This verifies the form of the eigenstates stated in \cref{fig:DressedHamiltonianSpace}.
	Conversely, given a time-independent $\hat{H}'$ separating in a $\hat{H}$ and sum of $\hat{H}_{\mathrm{aux},\alpha}$ (using these depend on $\br_1 \sigma_1,...,\br_N \sigma_N,p_1,...,p_M$ respectively $p_{\alpha,2},...,p_{\alpha,N}$), with proper exchange-symmetries, we may always pick all the \emph{relevant} eigenstates to be of this particular form.
	An initial state $\Psi'_0$ with (some of) the physical properties will thus also keep these under exact evolution by $\hat{H}'(t)$.
	Note that the ground state of $\hat{H}_{\mathrm{aux},\alpha}$ is just a product of harmonic oscillator ground states $(\tfrac{\omega_\alpha}{\pi})^{1/4} \exp(-\tfrac{\omega_\alpha}{2} p_{\alpha,k}^2)$, since this simple product has the $q_{\alpha,k} \leftrightarrow q_{\alpha,l}$ symmetry.
	However, this is special to the harmonic oscillator ground states, and excited $\chi_\alpha$ are thus as said beyond our scope.

	Note that $\hat{V}'(t)$ and $\hat{W}'$ do not individually possess the $\br_k \sigma_k \leftrightarrow \br_l \sigma_l$ and $q_{\alpha,k} \leftrightarrow q_{\alpha,l}$ exchange symmetries, only the $\bz_k \sigma_k \leftrightarrow \bz_l \sigma_l$ symmetry, in contrast to $\hat{V}(t)$ and $\hat{W}$.
	Further, they do not individually separate into terms of $\br_1 \sigma_1,...,\br_N \sigma_N,p_1,...,p_M$ and $p_{\alpha,2},...,p_{\alpha,N}$ as $\hat{H}'(t)$ does.
	This is another example of these properties rely on a fine balance between $\hat{V}'(t)$ and $\hat{W}'$, which is rather unique to $v'(\bz,t)$ and $w'(\bz,\bz')$ that correspond to a physical system.
	The only case where $\hat{V}'(t)$ and $\hat{W}'$ exhibit these physical properties individually is if all $\blambda_\alpha = 0$.

	For all $\blambda_\alpha = 0$, $\hat{H}'$ further separates in an electron and photon parts $\hat{H}' = \hat{H}_\mathrm{E} + \sum_{\alpha=1}^M \!\hat{H}'_{\mathrm{P}\!,\alpha}$, $\hat{H}'_{\mathrm{P}\!,\alpha} \!= \hat{H}_{p_\alpha} \!+ \hat{H}_{\mathrm{aux},\alpha}$.
	Here $\hat{H}_{p_\alpha} = - \tfrac{1}{2} \tfrac{\partial^2}{\partial p_\alpha^2} + \tfrac{1}{2} \omega_\alpha^2 (p_\alpha - \bar{p}_\alpha)^2 - \tfrac{1}{2} \omega_\alpha^2 \bar{p}_\alpha^2$ is a shifted harmonic oscillator shifted by $\bar{p}_\alpha = - \tfrac{\dot{J}_\alpha}{\omega_\alpha^3}$.
	The separating eigenstates of $\hat{H}'_{\mathrm{P},\alpha}$ take the form $\Phi'_{\mathrm{P},\alpha} = \varphi_{p_\alpha}(p_\alpha-\bar{p}_\alpha) \chi_\alpha$, where $\varphi_{p_\alpha}(p_\alpha)$ is a harmonic oscillator eigenstate, and $\chi_\alpha$ still is an eigenstate of $\hat{H}_\mathrm{aux,\alpha}$ with $q_{\alpha,k} \leftrightarrow q_{\alpha,l}$ symmetry (so both parts have this symmetry).
	In $q_{\alpha,k}$ coordinates, $\hat{H}'_{\mathrm{P},\alpha} \!=\! \sum_{k=1}^N \!\left[ - \tfrac{1}{2} \tfrac{\partial^2}{\partial q_{\alpha,k}^2} + \tfrac{1}{2} \omega_\alpha^2 (q_{\alpha,k} - \tfrac{\bar{p}_\alpha}{\sqrt{N}})^2 - \tfrac{1}{2N} \omega_\alpha^2 \bar{p}_\alpha^2 \right]$.
	Here the symmetric eigenstates are permanents $\Phi'_{q_\alpha}$ of $N$ orbitals, $\varphi'_{q_\alpha,k}(q_\alpha-\tfrac{\bar{p}_\alpha}{\sqrt{N}})$, where each orbital is a harmonic oscillator eigenstate that need not be the same for all $k$ (and we normalise all the permanents in this work to $1$).
	Many of these permanents are degenerate, namely if the orbital excitation levels add to the same total.
	To find the eigenstates that further separate in $\varphi_{p_\alpha}(p_\alpha - \bar{p}_\alpha) \chi_\alpha$ (i.e., to find $\Phi'_{\mathrm{P},\alpha}$ in $q_{\alpha,k}$ coordinates), we need to form linear combinations of these degenerate $\Phi'_{q_\alpha}$ permanents.
	It is beyond our scope to determine the coefficients of all these linear combinations, but it is easy to do for the low excitations we are mainly interested in.
	The ground state is a product (a single permanent of identical orbitals) $\prod_{k=1}^N \! \varphi'_{q_\alpha}\!(q_{\alpha,k} - \tfrac{\bar{p}_\alpha}{\sqrt{N}})$ of harmonic oscillator ground states (since then $\Phi'_{q_\alpha} = (\tfrac{\omega_\alpha}{\pi})^{N/4} \exp[-\tfrac{\omega_\alpha}{2} \sum_{k=1}^N (q_{\alpha,k}-\tfrac{\bar{p}_\alpha}{\sqrt{N}})^2]$ $= (\tfrac{\omega_\alpha}{\pi})^{N/4} \exp\{-\tfrac{\omega_\alpha}{2}[(p_\alpha - \bar{p}_\alpha)^2 + \sum_{k=2}^N p_{\alpha,k}^2]\} = \varphi_{p_\alpha}\chi_\alpha$).
	For low $\varphi_{p_\alpha}$ only a few coefficients need to be found.%
\footnote{%
	For example, for the second excited $\varphi_p$ and ground state $\chi$ we get
$\varphi_p \chi =\!
\tfrac{1}{\sqrt{2}} [2\omega(p-\bar{p})^2 - 1] (\tfrac{\omega}{\pi})^{N/4} \exp\{-\tfrac{1}{2}\omega[(p-\bar{p})^2+\sum_{k=2}^N p_k^2]\} =
\{\tfrac{1}{N}\sum_{k=1}^N \tfrac{1}{\sqrt{2}}[2\omega(q_k-\tfrac{\bar{p}}{\sqrt{N}})^2-1] + \tfrac{1}{\sqrt{2}N}\sum_{k \ne l = 1}^N \sqrt{2\omega}(q_k-\tfrac{\bar{p}}{\sqrt{N}}) \\ \sqrt{2\omega}(q_l-\tfrac{\bar{p}}{\sqrt{N}})\} (\tfrac{\omega}{\pi})^{N/4} \exp[-\tfrac{1}{2}\omega\sum_{k=1}^N (q_k - \tfrac{\bar{p}}{\sqrt{N}})^2]$.
	This is a linear combination of the two $\Phi'_{q_\alpha}$ permanents with excitations that add to $2$; one with all orbitals in their ground state except one second excited orbital, and one with two first excited orbitals.
	In general, the number of $\Phi'_{q_\alpha}$ permanents is given by the integer partition function $p(Q)$ of the total excitation level $Q$ (or is less for low $N$).
	This is easily managed for low excitations, and while $p(Q)$ grows as $\exp(\pi \!\sqrt{\scalebox{0.8}{$\frac{2}{3}$}Q})$ asymptotically the number of different $\varphi'_{q_\alpha,k}$ orbitals remains only $Q+1$, allowing further practical uses.
}

\subsection{Symmetries of Kohn-Sham Systems} \label{sec:SymmetriesOfKohnShamSystems}

	Recall from the introduction of this \namecref{sec:Symmetries} that the standard \ac{KS} Hamiltonian $\hat{H}'_\mathrm{KS}(t)$ takes the same form as the physical Hamiltonian $\hat{H}'(t)$ for $w(\br,\br')=\blambda_\alpha=0$, and therefore has all non-interacting physical properties.
	In \cref{sec:RelationshipWithStandardKS} we will thus show that $\hat{H}'_\mathrm{KS}(t) = \hat{H}_\mathrm{KS,E}(t) + \sum_{\alpha=1}^M \hat{H}'_{\mathrm{KS,P},\alpha}(t)$, just as for such non-interacting $\hat{H}'(t)$.
	The photon mode Hamiltonians $\hat{H}'_{\mathrm{KS,P},\alpha}(t)$ are the same as in the exact case for all $\blambda_\alpha = 0$, $\hat{H}'_{\mathrm{KS,P},\alpha}(t) = \hat{H}'_{\mathrm{P},\alpha}(t)$, only here $\bar{p}_\alpha(t) = - \tfrac{\dot{J}_\alpha(t)}{\omega_\alpha^3} + \tfrac{\blambda_\alpha \cdot \mathbf{R}(t)}{\omega_\alpha}$, where $\mathbf{R}(t)$ is the expectation value of the dipole operator $\hat{\mathbf{R}} = \sum_{k=1}^N \br_k$.
	They therefore have the same properties and eigenstates.
	The only case that the dressed $\hat{H}'_\mathrm{KS}(t)$ also has all these properties is if all $\blambda_\alpha = 0$, in which case the dressed and standard \ac{KS} descriptions usually coincide and reduce to purely electronic \ac{KS} as there is no electron-photon correlation.%
\footnote{%
	For a separating eigenstate $\Psi' = \Psi_\mathrm{E} \prod_{\alpha=1}^M \Phi'_{\mathrm{P}\!,\alpha}$ the dressed and standard \ac{KS} descriptions indeed coincide, as the corresponding standard \ac{KS} $\Phi' = \Phi_\mathrm{E} \prod_{\alpha=1}^M \Phi'_{\mathrm{P}\!,\alpha}$ shares the same density $n'(\bz)$.
	As standard \ac{KS} reduces to purely electronic, so does dressed \ac{KS}.
	For correlated $\Psi'_0$, the $n'(\bz,t)$, and hence the descriptions, differ though (and do not reduce to electronic \ac{KS}), even if all $\blambda_\alpha = 0$.
}
	In this case $v'_\mathrm{Mxc}(\bz,t)$ also reduces to the purely electronic Hartree exchange-correlation potential $v_\mathrm{Hxc}(\br,t)$, as the two-body part of $\hat{H}'(t)$ becomes purely electronic ($w'(\bz,\bz') = w(\br,\br')$).
	For $\blambda_\alpha \ne 0$, since $\hat{V}'(t)$ and $\hat{W}'$ individually only possess $\bz_k\sigma_k \leftrightarrow \bz_l\sigma_l$ symmetry, and the $n'(\bz,t)$ of $\Psi'(t)$ no longer separates, the dressed $\hat{H}'_\mathrm{KS}(t)$ also only has this symmetry, and the $\hat{H}'_\mathrm{KS}(t)$ and $v'_\mathrm{KS}(\bz,t)$ do not separate.%
\footnote{\label{foot:correlation}%
	It would be surprising if $v'_\mathrm{KS}(\bz,t) = v'(\bz,t) + v'_\mathrm{Mxc}(\bz,t)$ would separate or be on the form of \cref{DressedPhysicalPotential} (with the common $\blambda_\alpha = 0$ of \cref{DressedPhysicalPotential,DressedPhysicalInteraction} as $w'_\mathrm{KS}(\bz,\bz') = 0$) as $v'(\bz,t)$ only separates and is on this form for all $\blambda_\alpha = 0$.
	Likewise, it would be surprising if $\hat{V}'_\mathrm{KS}(t) = \hat{V}'(t) + \hat{V}'_\mathrm{Mxc}(t)$ had further properties than $\hat{V}'(t)$ has.
	To argue more strictly for most of these facts, we first recall that for the all $\blambda_\alpha = 0$ separating eigenstates $\Psi' = \Psi_\mathrm{E} \prod_{\alpha=1}^M \Phi'_{\mathrm{P}\!,\alpha}$, or the $\Phi' = \Phi_\mathrm{E} \prod_{\alpha=1}^M \Phi'_{\mathrm{P}\!,\alpha}$ of standard \ac{KS}, $n'(\bz)$ clearly separates.
	In contrast it does not for the $\blambda_\alpha \ne 0$ correlated $\Psi'$ eigenstates.
	The dressed $\Phi'$ that recreates a correlated $n'(\bz)$ therefore also has to be correlated (since a non-correlated $\Phi'$ would just give a non-correlated $n'(\bz)$).
	Now, since for all $\blambda_\alpha = 0$ we want dressed and standard \ac{KS} to coincide, $\Phi'$ and $\varphi'(\bz\sigma)$ separate in this case.
	Therefore, since the form of $\Phi'$ should not change for $\blambda_\alpha \ne 0$, only the $\varphi'(\bz\sigma)$, the correlation of $\Phi'$ must come from the $\varphi'(\bz\sigma)$ get correlated.
	In turn this implies $v'_\mathrm{KS}(\bz,t)$ must get correlated.
	Note that the separation or correlation of $v'_\mathrm{KS}(\bz,t)$ is also directly related with the exchange symmetries.
	For example, for two electrons and one mode, $\hat{V}'_\mathrm{KS}(t) = v'_\mathrm{KS}(\br_1,q_1,t) + v'_\mathrm{KS}(\br_2,q_2,t)$.
	The only case in which $\hat{V}'_\mathrm{KS}(t)$ is symmetric under exchange of $\br_1 \sigma_1$ and $\br_2 \sigma_2$ or $q_1$ and $q_2$ is when $v'_\mathrm{KS}(\br,q,t)$ separates into $v_\mathrm{KS,\br}(\br,t)+v'_\mathrm{KS,q}(q,t)$.
	This also implies that no dressed scheme can have correlated orbitals, yet keep the exchange symmetries, as used in \cref{foot:premise}.
	For the $\hat{V}'_\mathrm{KS}(t)$ in the example to separate in two terms of $\br_1 \sigma_1,\br_2 \sigma_2,p_1$ and $p_{1,2}$ further requires a suitable $v'_\mathrm{KS,q}(q,t)$.	
	Although we do not show it, it is thus highly unlikely that $\hat{V}'_\mathrm{KS}(t)$ separates like this.
	The physical $v'(\bz,t)$ and $w'(\bz,\bz')$ of \cref{DressedPhysicalPotential,DressedPhysicalInteraction} are thus very special to have the discussed physical properties.
}
	In dressed \ac{KS}, approximate $v'_\mathrm{Mxc}(\bz,t)$ should therefore also only separate if all $\blambda_\alpha \!= 0$.

	Given a time-independent standard \ac{KS} Hamiltonian $\hat{H}'_\mathrm{KS}$, or a dressed with all $\blambda_\alpha = 0$, we may thus compute the $\br_k \sigma_k \leftrightarrow \br_l \sigma_l$ anti-symmetric eigenstates $\Phi_\mathrm{E}$ of $\hat{H}_\mathrm{KS,E}$.
	These take the form of a single Slater determinant $\Phi_{\br\sigma}$ of $\varphi_{\br \sigma,k}(\br \sigma)$ orbitals (or a linear combination of a few such) as the \ac{KS} electrons do not interact.
	We may again further compute the separating, $q_{\alpha_{\!},k} \!\leftrightarrow\! q_{\alpha_{\!},l}$ symmetric eigenstates $\Phi'_{\mathrm{P},\alpha}$ of the $\hat{H}'_{\mathrm{KS,P},\alpha}$.
	We may then combine the $\Phi_\mathrm{E}$ and $\Phi'_{\mathrm{P},\alpha}$ to form the eigenstates $\Phi' = \Phi_\mathrm{E} \prod_{\alpha=1}^M \Phi'_{\mathrm{P},\alpha}$ of $\hat{H}'_\mathrm{KS}$, which satisfy all the non-interacting physical properties.
	In $\bz \sigma$ coordinates the combined $\Phi'$ ground state is a single Slater determinant	 of the combined orbitals $\,\varphi'_k(\bz \sigma)\, = \varphi_{\br \sigma,k}(\br \sigma) \prod_{\alpha=1}^M \varphi'_{q_\alpha}(q_\alpha - \tfrac{\bar{p}_\alpha}{\sqrt{N}})$ (or a linear combination of such determinants in case $\Phi_\mathrm{E}$ is a linear combination).
	This follows as $\prod_{\alpha=1}^M \Phi'_{\mathrm{P},\alpha} = \prod_{\alpha=1}^M \prod_{k=1}^N \varphi'_{q_\alpha}(q_{\alpha,k}-\tfrac{\bar{p}_\alpha}{\sqrt{N}})$ in this case (as showed earlier), which is a common part in all terms of the Slater determinant for the photon part.
	In ground-state \ac{KS} this is all we need.
	In general though, the eigenstates are linear combinations of $M$-fold sums $C \sum_{\tau^1} ... \sum_{\tau^M} \Phi'_{\tau^1,...,\tau^M}$ of Slater determinants $\Phi'_{\tau^1,...,\tau^M}$ (note that we consider a normalisation constant $C$ a part of each $M$-fold sum to keep each $M$-fold sum normalised).		
	This follows since $\Phi_\mathrm{E}$ and each $\Phi'_{\mathrm{P},\alpha}$ is a linear combination of determinants $\Phi_{\br \sigma}$ respectively permanents $\Phi'_{q_\alpha}$, and each determinant-permanents product $\Phi_{\br\sigma} \prod_{\alpha=1}^M \!\Phi'_{q_\alpha}$ equals such an $M$-fold sum.
	The $N$ orbitals of each of the Slater determinants $\Phi'_{\tau^1,...,\tau^M}$ are given by $\varphi'_{\tau^1\!,...,\tau^M\!,k}(\bz \sigma) \\ = \varphi_{\br\sigma,k}(\br\sigma) \prod_{\alpha=1}^M \varphi'_{q_\alpha,\tau_k^\alpha}(q_\alpha - \tfrac{\bar{p}_\alpha}{\sqrt{N}})$.
	Each of the $\tau^\alpha$ sums in the $M$-fold sum then runs over all permutations $\tau^\alpha$ of the sequence $(1,...,N)$, and acts as a symmetrizer for the $q_{\alpha,k} \leftrightarrow q_{\alpha,l}$ exchange symmetry for the given $\alpha$ since it sums over all permutations of the $\varphi'_{q_\alpha,k}(q_\alpha - \tfrac{\bar{p}_\alpha}{\sqrt{N}})$ orbitals.
	Note that the $M$-fold sums usually simplify greatly due to identical orbitals in the $\Phi'_{q_\alpha}$ permanents.
	For example, in the special case that all orbitals in each permanent $\Phi'_{q_\alpha}$ are the same, all the $\Phi'_{\tau^1,...,\tau^M}$ become identical, and the $M$-fold sum reduces to a single determinant of $\varphi'_{k}(\bz\sigma) =\\ \varphi_{\br\sigma\!,k}(\br\sigma) \!\prod_{\alpha=1}^M \!\varphi'_{q_\alpha}\!(q_{\alpha\!}-_{\!}\tfrac{\bar{p}_\alpha}{\sqrt{N}}_{\!})$ orbitals as for ground states.%
\footnote{%
	The actual number of determinants of $\Phi'$ still grows fast with the excitation levels, but at least for few modes and low excitations the number of different $\bz\sigma$-orbitals remains feasible.
	This allows all practical uses expressible directly in terms of the $\bz\sigma$-orbitals in an efficient form.
	This includes propagation of the $\bz\sigma$-orbitals, but also say density computations, as we may compute the $n'(\bz)$ of an $M$-fold sum $\Phi_{\br\sigma} \prod_{\alpha=1}^M \Phi'_{q_\alpha}$ directly from the $\bz\sigma$-orbitals by $n'(\bz) \!=\!\! \left[ \sum_{k=1}^N \!|\varphi_{\br\sigma_{\!},k}(\br\sigma)|^2 \right] \!\prod_{\alpha=1\!}^M \!\left[ \frac{1}{N} \!\sum_{k_\alpha=1}^N \!|\varphi'_{q_{\alpha_{\!}},k_{\alpha\!}}\!(q_\alpha\!-\!\tfrac{\bar{p}_\alpha}{\sqrt{N}})|^2 \right]$ $=\!\frac{1}{N^M} \sum_{k,k_1,...,k_M=1}^N \!\left| \varphi_{\br\sigma_{\!},k}(\br\sigma) \prod_{\alpha=1}^M \varphi'_{q_\alpha,k_\alpha}\!(q_\alpha\!-\!\tfrac{\bar{p}_\alpha}{\sqrt{N}}) \right|^2\!\!$, which usually simplifies a lot further due to identical orbitals.
	The $n'(\bz)$ of the full $\Phi'$ is then a linear combination of the densities of the $N_{\br\sigma} p(Q_1)...p(Q_M)$ $M$-fold sums of $\Phi'$, where $N_{\br\sigma}$ is the number of $\Phi_{\br\sigma}$ in $\Phi_E$ and $p(Q_\alpha)$ is the integer partition function of the total excitation level of mode $\alpha$.
	This already directly allows for few modes and low excitations, and the many identical orbitals also allow further simplifications here, to allow higher excitations.
}
	Also note that if we to compute a ground state $\Phi'$ use the full $v'_\mathrm{KS}(\bz)$ to find the $\varphi'_k(\bz\sigma)$, instead of as above use the separation to find $\varphi_{\br\sigma,k}(\br\sigma)$ and $\varphi'_{q_\alpha}(q_\alpha)$ individually, we must still require the $\varphi_{\br\sigma\!,k}(\br\sigma)$ part of all $\varphi'_k(\bz\sigma)$ (of each $\Phi'_{\bz\sigma}$) differ (as the orbitals of each $\Phi'_\mathrm{\br\sigma}$ must still differ).
	The orbitals with lowest energy then all have the same ground state photon part and we get the $\varphi'_k(\bz\sigma)$ of before.
	For excited $\Phi'$ one has to identify all eigenorbitals $\varphi'(\bz \sigma)$ that correspond to a $\varphi_{\br\sigma,k}(\br\sigma) \prod_{\alpha=1}^M \varphi'_{q_\alpha,\tau_k^\alpha}(q_\alpha - \tfrac{\bar{p}_\alpha}{\sqrt{N}})$ of the given $\Phi'$.
	Unlike the individual, degenerate, $\Phi'_{\tau^1,...,\tau^M}$ or if not using the right orbitals (e.g., for a ground state), the proper $\Phi'$ as said have all the non-interacting physical properties (instead of only $\bz_k \sigma_k \leftrightarrow \bz_l \sigma_l$ anti-symmetry).
	Especially, they yield a different expression for $n'(\bz)$ in terms of the orbitals where $n'(\bz)$ always separates.
	This is another strong argument to use these $\Phi'$ in dressed \ac{KS} for all $\blambda_{\alpha\!} = 0$, as the exact physical eigenstates $\Psi'$ we are to mimic in this case also exhibit all these properties and separating $n'(\bz)$.
	In fact most other $\Phi'$ do not lead to a valid \ac{KS} system as they cannot reproduce the separating $n'(\bz)$ they here are to recreate (of the all $\blambda_\alpha = 0$ ground state $\Psi'\!$, or to be general of any all $\blambda_\alpha \!= 0$ eigenstate $\Psi'$).

	To generalise this to non-zero $\blambda_\alpha$ also for dressed $\hat{H}'_\mathrm{KS}$, we may use the same linear combinations of determinants for the $\Phi'$ as we do in standard \ac{KS}.
	For example, we may for the $\Phi'$ ground state still use a single (or simple linear combination of) determinant(s) of $\varphi'_k(\bz\sigma)$ orbitals.
	Only, as the $v'_\mathrm{KS}(\bz)$ no longer separates neither do the orbitals, and $\Phi'$ only has $\bz_k \sigma_k \leftrightarrow \bz_l \sigma_l$ anti-symmetry.
	To recover the right form if we set all $\blambda_\alpha \!= 0$, we should again select the orbitals with lowest energy that further have different electron parts, though as the orbitals no longer perfectly separate, we need to refine what exactly we mean by this.
	The optimal way to do this is beyond our scope, but it is usually easy as long as the interaction is not too strong.
	For excited $\Phi'$ one has to identify all $\varphi'(\bz \sigma)$ that for $\blambda_\alpha = 0$ correspond to a $\varphi_{\br\sigma,k}(\br\sigma) \prod_{\alpha=1}^M \varphi'_{q_\alpha,\tau_k^\alpha}(q_\alpha - \tfrac{\bar{p}_\alpha}{\sqrt{N}})$ of $\Phi'$.
	Using the $\Phi'$ linear combinations of $M$-fold sums with these orbitals we then recover the right $\Phi'$ for all $\blambda_\alpha = 0$.
	As pointed out in the introduction to this \namecref{sec:Symmetries} these $\Phi'$ are usually not eigenstates of $\hat{H}'_\mathrm{KS}$ for $\blambda_\alpha \ne 0$ though, except in single determinant cases (or if the determinants only differ in their spin), as the correlation of the $\varphi'(\bz\sigma)$ orbitals usually lifts the degeneracy of the determinants.
	If we insist on using eigenstates of $\hat{H}'_\mathrm{KS}$ in dressed \ac{KS}, these are the single determinants of the $\Phi'$ we propose, or linear combinations of such determinants that remain degenerate.
	Both options differ significantly from the full $\Phi'$ we propose, and so break the important equivalences for all $\blambda_\alpha\!=\!0$ discussed in the introduction to this \namecref{sec:Symmetries}.
	We would thus have to mimic states of one form by states of a very different form for all $\blambda_\alpha=0$, and therefore also for other $\blambda_\alpha$.
	We would also run into the issues discussed at the end of the last paragraph, which would also persist for other $\blambda_\alpha$ (even if some $v'_\mathrm{KS}(\bz)$ is able to recreate $n'(\bz)$ if all $\blambda_\alpha$ are non-zero it would not give anything sensible).
	It is beyond our scope to analyse the precise consequences of not using eigenstates for $\Phi'$, but it seems a lessor issue.

	For propagation, $\Phi'(t)$ keeps its initial $\bz\sigma$-orbital form if it is a linear combination of determinants (like above), so it stays $\bz_k \sigma_k \leftrightarrow \bz_l \sigma_l$ anti-symmetric and numerically efficient.\!
	The initial state $\Phi'_0$ may have further properties, but these are only kept under evolution using a standard $\hat{H}'_\mathrm{KS}(t)$, or a dressed $\hat{H}'_\mathrm{KS}(t)$ if all $\blambda_\alpha = 0$.
	For example, our examples in the main text are special in the initial states are prepared outside the cavity for $\blambda_\alpha = \dot{J}_\alpha(t) = 0$, so the $\Phi'_0$ have all the non-interacting physical properties, but the $\Phi'(t)$ only keep the $\bz_k \sigma_k \leftrightarrow \bz_l \sigma_l$ anti-symmetry.

	Finally, note that even if $\Phi'(t)$ only has $\bz_k \sigma_k \leftrightarrow \bz_l \sigma_l$ anti-symmetry, many derived physical quantities such as all electron $n$-body densities and density matrices recover appropriate exchange symmetries due to this symmetry.
	For example, take a spin-independent expectation value $O(\br_1,...,\br_n,\{p_\beta\},t) = \braket<\Phi'(t)|\hat{O}(\br_1,...,\br_n,\{p_\beta\})|\Phi'(t)>$.
	In general it may depend on $0 \le n \le N$ spatial coordinates and a subset $\{p_\beta\}$ of the photon coordinates $p_\alpha$.
	To evaluate it, one generally has to perform a coordinate transformation:
	Typically, $\Phi'(t)$ is given in terms of the $q_{\alpha,k}$-coordinates, and one transforms the $p_\beta$ of $\hat{O}$ to $q_{\beta,k}$, though one may also transform $\Phi'(t)$ to $p_{\alpha,k}$-coordinates.
	Since the RHS is typically a spin-summed integral of a $\bz_k \sigma_k \leftrightarrow \bz_l \sigma_l$ symmetric integrand, also the LHS has this symmetry.
	For the LHS this is equivalent with $\br_k \leftrightarrow \br_l$ symmetry, since all the $p_\beta$ are symmetric under exchange of $q_{\beta,k}$ and $q_{\beta,l}$, so this part of the $\bz_k \sigma_k \leftrightarrow \bz_l \sigma_l$ exchange has no effect, and the two kinds of exchange are identical.
	Note the argument generalises trivially to spin-dependent operators; we only restricted us for notational simplicity.%
\footnote{\label{foot:unique}%
	Also note that $n(\br,t) = N \sum_{\sigma_1,...,\sigma_N} \int \diff^{3(N-1)} r \, \diff^{MN} q \, |\Phi'(t)|^2$ yields the same no matter which $\br_k$ one does not integrate over, even if $\Phi'(t)$ has no $\br_k\sigma_k \leftrightarrow \br_l\sigma_l$ anti-symmetry.
	This follows as $\sum_{\sigma_1,...,\sigma_N} \int \diff^{MN} q \, |\Phi'(t)|^2$ has $\br_k \leftrightarrow \br_l$ symmetry by a similar argument as above.
	The asymmetric definition, the symmetric $n(\br,t) = \sum_{k=1}^N \braket<\Phi'(t)|\delta(\br-\br_k)|\Phi'(t)>$ and $n(\br,t) = \int \!\diff^M q \, n'(\bz,t)$ therefore all give the same $n(\br,t)$.
	Similarly, for example $p_\alpha(t)$ and $\rho_2(\br,\br',t)$ remain uniquely defined.
	For approximate $\Psi'(t)$, if $\Psi'(t) = \Psi(t)\chi(t)$, $n(\br,t)$, $p_\alpha(t)$ and $\rho_2(\br,\br',t)$ are for example further the same whether computed using $\Psi'(t)$ or $\Psi(t)$, even if $\Psi'(t)$ lacks $\br_k\sigma_k \leftrightarrow \br_l\sigma_l$ anti-symmetry or $q_{\alpha,k} \leftrightarrow q_{\alpha,l}$ symmetry.
}\!\!
	In general it is thus only internal quantities like $\rho'_2(\bz,\bz'\!,t)$ that lack the $\br_k \leftrightarrow \br_l$ (anti)-symmetry.
	Also, even if $\Phi'(t)$ does not separate in a $\Phi(t)\chi(t)$, all $O(\br_1,...,\br_n,\{p_\beta\},t)$ expectation values still only depend on the coordinates of $\Phi(t)$.

\section{Equations of Motion and Forces} \label{sec:EOM}

	In this \namecref{sec:EOM} we first present the equations of motion for $n(\br,t)$ and $p_\alpha(t)$ in the physical coordinates, to obtain the physical forces we want to capture with the dressed \ac{KS} scheme.
	We then present the equation of motion for $n'(\bz,t)$ in the auxiliary coordinates for general $v'(\bz,t)$ and $w'(\bz,\bz')$, which has the very same structure as the usual divergence of local-forces equation.
	We then study the special case where $v'(\bz,t)$ and $w'(\bz,\bz')$ correspond to a physical system characterised by a $v(\br,t)$, $w(\br,\br')$ and $\dot{J}_\alpha(t)$.
	We again present the equations of motion for $n(\br,t)$ and $p_\alpha(t)$, but this time in terms of the quantities of the dressed system.
	This then allows us to see how the dressed system produces the same forces as the physical, and to establish \textit{exact relations} between the dressed and physical forces.
	Finally, we again present the equations of motion for $n'(\bz,t)$, $n(\br,t)$ and $p_\alpha(t)$, but this time in terms of the quantities of the dressed \ac{KS} system, to see how this system also produces the different forces.

\subsection{Physical Equations of Motion}

	We consider the physical electron-photon Hamiltonian
\begin{align} \label{Hamiltonian}
\hat{H}(t) &= \sum_{k=1}^N \left[ -\tfrac{1}{2} \nabla_{\br_k}^2 + v(\br_k, t) \right] + \tfrac{1}{2} \sum_{k \neq l} w(\br_k,\br_l) \\
&\!\!\!\!\!+ \sum_{\alpha=1}^M \bigg[ - \tfrac{1}{2} \tfrac{\partial^2}{\partial p_\alpha^2} + \tfrac{1}{2} \Big( \omega_\alpha p_\alpha - \blambda_\alpha \cdot \sum_{k=1}^N \br_k \Big)^2 \!\! + \tfrac{\dot{J}_\alpha(t)}{ \omega_\alpha} p_\alpha \bigg] . \nonumber
\end{align}

	The expectation values $n(\br,t) = \braket<\Psi(t)|\hat{n}(\br)|\Psi(t)>$ and $\bj(\br,t) = \braket<\Psi(t)|\hat{\bj}(\br)|\Psi(t)>$ of the physical density and current operators $\hat{n}(\br) = \sum_{k=1}^N \delta(\br-\br_k)$ and $\hat{\bj}(\br) = \tfrac{1}{2\imagi} \sum_{k=1}^N \,[\delta(\br - \br_k) \overrightarrow{\nabla}_{\br_k} - \overleftarrow{\nabla}_{\br_k} \delta(\br - \br_k)]$ then obey the continuity equation
\begin{align*}
\tfrac{\partial}{\partial t} n(\br,t) = - \nabla_\br \cdot \bj(\br,t) .
\end{align*}
	Taking the second time-derivative, we then arrive at the divergence of local-force equation \cite{ruggenthaler2015ex}
\begin{align*}
\tfrac{\partial^2}{\partial t^2} &n(\br,t) = \nabla_\br \cdot \left[ n(\br,t) \nabla_\br v(\br,t) \right] \\
& - \nabla_\br \cdot [\mathbf{Q}_\mathrm{kin}(\br,t) + \mathbf{Q}_\mathrm{int}(\br,t) + \bF_\mathrm{dip}(\br,t) + \bF_\mathrm{lin}(\br,t)] ,
\end{align*}
where $\mathbf{Q}_\mathrm{kin}(\br,t) = \imagi \braket<\Psi(t)|[\hat{T},\hat{\bj}(\br)]|\Psi(t)>$ and $\mathbf{Q}_\mathrm{int}(\br,t) = \imagi \braket<\Psi(t)|[\hat{W},\hat{\bj}(\br)]|\Psi(t)> = -2 \int \diff \br' \rho_2(\br,\br',t) \nabla_\br w(\br,\br')$ are the physical momentum-stress and interaction-stress forces, and,
\begin{align*}
\bF_\mathrm{dip}(\br,t) &= - \sum_{\alpha=1}^M \blambda_\alpha \braket<\Psi(t)| (\blambda_\alpha \cdot \sum_{k=1}^N \br_k ) \hat{n}(\br) |\Psi(t)> = \\
&\hspace{-14mm} \underbrace{- \sum_{\alpha=1}^M \blambda_\alpha n(\br t) (\blambda_\alpha \cdot \br)}_{\bF_\mathrm{dip}^{(1)}(\br,t)}
\underbrace{- 2 \sum_{\alpha=1}^M \blambda_\alpha \int \diff \br' \rho_2(\br,\br',t) (\blambda_\alpha \cdot \br')}_{\bF_\mathrm{dip}^{(2)}(\br,t)} , \\
\bF_\mathrm{lin}(\br,t) &= \sum_{\alpha=1}^M \blambda_\alpha \braket<\Psi(t)| \omega_\alpha p _\alpha \hat{n}(\br) |\Psi(t)> ,
\end{align*}
are the forces the photons exert on the electron density \cite{tokatly2013}.
	Here $\bF_\mathrm{lin}(\br,t)$ is due to the bilinear coupling between the displacement field and the electrons, while $\bF_\mathrm{dip}(\br,t)$ is due to the dipole self-interaction, and balances the bilinear coupling such that the resulting Hamiltonian stays bounded from below, i.e., allows for a ground state \cite{rokaj2017}.

	The physical displacement coordinates similarly satisfy the mode-resolved Maxwell equations \cite{tokatly2013}
\begin{align*}
\tfrac{\partial^2}{\partial t^2} p_\alpha(t) = - \omega_\alpha^2 p_\alpha(t) + \omega_\alpha \blambda_\alpha \cdot \mathbf{R}(t) - \tfrac{\dot{J}_\alpha(t)}{\omega_\alpha} ,
\end{align*}
where $\mathbf{R}(t) = \int \diff^3 r \, \br n(\br,t)$ is the total dipole.

\subsection{Dressed Equations of Motion}

	We consider the dressed electron-photon Hamiltonian
\begin{align*}
\hat{H}'(t) &= \sum_{k=1}^N \left[ -\tfrac{1}{2} \nabla_{\bz_k}^2 + v'(\bz_k,t) \right] + \tfrac{1}{2} \sum_{k \neq l} w'(\bz_k,\bz_l) .
\end{align*}

	The expectation values $n'(\bz,t) = \braket<\Psi'(t)|\hat{n}'(\bz)|\Psi'(t)>$ and $\bj'(\bz,t) = \braket<\Psi'(t)|\hat{\bj}'(\bz)|\Psi'(t)>$ of the dressed density and current operators $\hat{n}'(\bz) = \sum_{k=1}^N \delta(\bz-\bz_k)$ and $\hat{\bj}'(\bz) = \tfrac{1}{2 \imagi} \sum_{k=1}^N [\delta(\bz - \bz_k) \overrightarrow{\nabla}_{\bz_k} - \overleftarrow{\nabla}_{\bz_k} \delta(\bz - \bz_k)]$ then obey the continuity equation
\begin{align*}
\tfrac{\partial}{\partial t} n'(\bz,t) = - \nabla_\bz \cdot \bj'(\bz,t) .
\end{align*}
	Taking the second time-derivative, we then arrive at
\begin{align} \label{DressedEOM}
\tfrac{\partial^2}{\partial t^2} n'(\bz,t) &= \nabla_\bz \cdot \left[ n'(\bz,t) \nabla_\bz v'(\bz,t) \right] \\
&- \nabla_\bz \cdot [\mathbf{Q}'_\mathrm{kin}(\bz,t) + \mathbf{Q}'_\mathrm{int}(\bz,t)] ,\nonumber
\end{align}
where $\mathbf{Q}'_\mathrm{kin}(\bz,t) = \imagi \braket<\Psi'(t)|[\hat{T}',\hat{\bj}'(\bz)]|\Psi'(t)>$ and $\mathbf{Q}'_\mathrm{int}(\bz,t) = \imagi \braket<\Psi'(t)|[\hat{W}',\hat{\bj}'(\bz)]|\Psi'(t)> = -2 \int \diff \bz' \rho'_2(\bz,\bz',t) \nabla_\bz w'(\bz,\bz')$ are the dressed momentum-stress and interaction-stress forces \cite{tokatly2005}.

\subsection{Physical Dressed Equations of Motion} \label{sec:PhysicalDressedEquationsOfMotion}

	Substituting the expressions for $v'(\bz,t)$ and $w'(\bz,\bz')$ in \cref{DressedEOM} for a physical system, we obtain the equation of motion for $n'(\bz,t)$ in terms of $v(\br,t)$, $w(\br,\br')$ and $\dot{J}_\alpha(t)$.
	This special case applies only for $v'(\bz,t)$ and $w'(\bz,\bz')$ that correspond to a physical system.

	Integrating the resulting equation over all $q_\alpha$ coordinates, using that the integral of a divergence vanishes for a closed system, we again find the equation of motion for $n(\br,t)$, but this time expressed in terms of the quantities of the dressed system, i.e.,
\begin{align*}
\tfrac{\partial^2}{\partial t^2} &n(\br,t) = \nabla_\br \cdot \left[ n(\br,t) \nabla_\br v(\br,t) \right] \\
&- \nabla_\br \cdot \left[ \mathbf{Q}_\mathrm{kin}^\mathrm{d}(\br,t) + \mathbf{Q}_\mathrm{int}^\mathrm{d}(\br,t) + \bF_\mathrm{dip}^\mathrm{d}(\br,t) + \bF_\mathrm{lin}^\mathrm{d}(\br,t) \right] .
\end{align*}
	Here the dressed stress and photon-matter forces are
\begin{align*}
\mathbf{Q}_\mathrm{kin}^\mathrm{d}(\br,t) &= \imagi \braket<\Psi'(t)|[\hat{T},\hat{\bj}(\br)]|\Psi'(t)> , \\
\mathbf{Q}_\mathrm{int}^\mathrm{d}(\br,t) &= 
-2 \int \diff \br' \, \rho_2(\br,\br',t) \nabla_\br w(\br,\br') \\
&+ (N-1) \bF_\mathrm{lin}^\mathrm{d}(\br,t) \\
&-2 \sum_{\alpha=1}^M \blambda_\alpha \int \diff \br' \rho_2(\br,\br',t) (\blambda_\alpha \cdot \br') , \\
\bF_\mathrm{dip}^\mathrm{d}(\br,t) &= - \sum_{\alpha=1}^M \blambda_\alpha n(\br,t) (\blambda_\alpha \cdot \br) , \\
\bF_\mathrm{lin}^\mathrm{d}(\br,t) &= \sum_{\alpha=1}^M \blambda_\alpha \int \diff^M q \tfrac{\omega_\alpha q_\alpha}{\sqrt{N}} n'(\bz,t) ,
\end{align*}
where we note the 2\textsuperscript{nd} term $(N-1) \bF_\mathrm{lin}^\mathrm{d}(\br,t)$ of $\mathbf{Q}_\mathrm{int}^\mathrm{d}(\br,t)$ relies on the $q_{\alpha,k} \leftrightarrow q_{\alpha,l}$ exchange symmetry to establish that $2 \int \diff \br' \diff^M q \, \rho'_2(\bz,\bz',t) = (N-1) n'(\br,q'_1,...,q'_M,t)$.
	For later reference when we introduce approximations, we note this is the only property the above expressions rely on specific to the subset of physical $v'(\br,t)$ and $w'(\bz,\bz')$.%
\footnote{%
	Remember from \cref{foot:unique} that $n(\br,t)$, $\rho_2(\br,\br',t)$ and $p_\alpha(t)$ are uniquely defined even if $\Psi'(t)$ only has $\bz_k\sigma_k \leftrightarrow \bz_l\sigma_l$ symmetry, so we need no further properties to use these here.
}
	Breaking this symmetry by, for example, adding a $\delta v'\!(\bz,\!t)$ without this symmetry, one has to replace the given term by $2 \sum_{\alpha=1}^M \blambda_\alpha \int \diff^M q \diff^{3+M} z' \, \tfrac{\omega_\alpha q'_\alpha}{\sqrt{N}} \rho'_2(\bz,\bz',t)$, while all the other terms above remain unchanged.

	Multiplying the equation of motion for $n'(\bz,t)$ by $\tfrac{q_\alpha}{\sqrt{N}}$, and integrating this over all $\bz$ coordinates using \cref{pAlpha}, we again find the mode-resolved Maxwell equations,
\begin{align*}
\tfrac{\partial^2}{\partial t^2} p_\alpha(t) = - \omega_\alpha^2 p_\alpha(t) + \omega_\alpha \blambda_\alpha \cdot \mathbf{R}(t) - \tfrac{\dot{J}_\alpha(t)}{\omega_\alpha} ,
\end{align*}
with a contribution from $v'(\bz,t)$ of
\begin{align*}
- \omega_\alpha^2 p_\alpha(t) + \tfrac{\omega_\alpha}{N} \blambda_\alpha \cdot \mathbf{R}(t) - \tfrac{\dot{J}_\alpha(t)}{\omega_\alpha} ,
\end{align*}
and a contribution from $w'(\bz,\bz')$ of
\begin{align*}
\tfrac{(N-1)\omega_\alpha}{N} \blambda_\alpha \cdot \mathbf{R}(t) .
\end{align*}
	In the physical coordinates the mode-resolved Maxwell equations instead originated from the second line of the physical Hamiltonian \labelcref{Hamiltonian}.

\subsection{Comparison of Force Terms: Exact Relations}

	The equations of motion for $n(\br,t)$ and $p_\alpha(t)$ in terms of the quantities of the physical and exact dressed system must of course be the same, as trivial to confirm for $p_\alpha(t)$.
	For the electron density $n(\br,t)$ this implies that
\begin{align*}
\mathbf{Q}_\mathrm{kin}^\mathrm{d}(\br,t) + \mathbf{Q}_\mathrm{int}^\mathrm{d}(\br,t) +
\bF_\mathrm{dip}^\mathrm{d}(\br,t) + \bF_\mathrm{lin}^\mathrm{d}(\br,t) = \\
\mathbf{Q}_\mathrm{kin}(\br,t) + \mathbf{Q}_\mathrm{int}(\br,t) +
\bF_\mathrm{dip}(\br,t) + \bF_\mathrm{lin}(\br,t) .
\end{align*}
	This we can also easily confirm by using that for a dressed system corresponding to a physical one $\Psi'(t)=\Psi(t)\chi(t)$, which together with the $q_{\alpha,k} \leftrightarrow q_{\alpha,l}$ exchange symmetry allows us to establish the following exact relations%
\footnote{%
	Note that the physical force expressions also apply even for some non-physical $\Psi'(t)$.
	In fact the $\mathbf{Q}_\mathrm{int}(\br,t)$, $\bF_\mathrm{dip}^{(1)}(\br,t)$ and $\bF_\mathrm{dip}^{(2)}(\br,t)$ expressions are useful for any $\Psi'(t)$.
	$\mathbf{Q}_\mathrm{kin}(\br,t)$ and $\bF_\mathrm{lin}(\br,t)$ are only defined given a $\Psi(t)$, so for these we need $\Psi'(t) = \Psi(t)\chi(t)$, in which case one can also compute $n(\br,t)$ and $\rho_2(\br,\br',t)$ by $\Psi(t)$ instead of $\Psi'(t)$.
	We further assumed $q_{\alpha,k} \leftrightarrow q_{\alpha,l}$ symmetry to obtain the last relation and the second term of $\mathbf{Q}_\mathrm{int}^\mathrm{d}(\br,t)$, though it is possible that $\Psi'(t) = \Psi(t) \prod_{\alpha=1}^M \chi_\alpha(t)$ would suffice instead.
	Besides these assumptions, all we used to show the four relations are the definitions of $\hat{n}'(\bz)$, $p_\alpha$ and $\hat{n}(\br)$ used for the last relation.
}
\begin{align*}
\mathbf{Q}_\mathrm{kin}^\mathrm{d}(\br,t) &= \mathbf{Q}_\mathrm{kin}(\br,t) , \\
\mathbf{Q}_\mathrm{int}^\mathrm{d}(\br,t) &= \mathbf{Q}_\mathrm{int}(\br,t)
+ \tfrac{N-1}{N} \bF_\mathrm{lin}(\br,t) + \bF_\mathrm{dip}^{(2)}(\br,t) , \\
\bF_\mathrm{dip}^\mathrm{d}(\br,t) &= \bF_\mathrm{dip}^{(1)}(\br,t) , \\
\bF_\mathrm{lin}^\mathrm{d}(\br,t) &= \tfrac{1}{N} \bF_\mathrm{lin}(\br,t) .
\end{align*}
	These relations also provide insight into how the forces get recast in the dressed system.
	For example, they show that the kinetic term is the same in both cases, and that $\bF_\mathrm{lin}^\mathrm{d}(\br,t)$ gives $\tfrac{1}{N} \bF_\mathrm{lin}(\br,t)$, while $\mathbf{Q}_\mathrm{int}^\mathrm{d}(\br,t)$ yields the rest.
	Further, $\bF_\mathrm{dip}(\br,t)$ comes from the $(\blambda_\alpha \cdot \hat{\mathbf{R}})^2$ term in $\hat{H}(t)$, which can be split into a one-body and two-body part, i.e., a potential and interaction.
	The one-body part leads to $\bF_\mathrm{dip}^{(1)}(\br,t)$ both in the physical and the dressed system, while the two-body part provides the remaining $\bF_\mathrm{dip}^{(2)}(\br,t)$,\!\! so the dipole self-interaction is indeed treated the same in the two formulations (also in case of a \ac{KS} calculation).

\subsection{Dressed Kohn-Sham Equations of Motion}

	We consider the dressed \ac{KS} Hamiltonian
\begin{align*}
\hat{H}'_\mathrm{KS}(t) = \sum_{k=1}^N \left[ -\tfrac{1}{2} \nabla_{\bz_k}^2 + v'(\bz_k,t) + v_\mathrm{Mxc}'(\bz_k,t) \right] ,
\end{align*}
of a non-interacting auxiliary \ac{KS} system, that if using a self-consistent time-propagation by design reproduces the physical dressed density $n'[\Psi'_0,v']$ of propagating $\Psi'_0$ by $v'(\bz,t)$.
	To ensure this, the \ac{Mxc} potential $v'_\mathrm{Mxc}[\Psi'_0,\Phi'_0,n']$\!\! is defined as the potential difference $v'_s[\Phi'_0,n']-v[\Psi'_0,n']$ between a non-interacting and interacting system with the same dressed density $n'(\bz,t)$, since the \ac{KS} potential $v'_\mathrm{KS}[\Psi'_0,\Phi'_0,n',v'] \!=\! v' + v'_\mathrm{Mxc}[\Psi'_0,\Phi'_0,n']$ for the desired dressed density $n'[\Psi'_0,v']$ then reduces to the unique non-interacting potential $v'_s[\Phi'_0,n'[\Psi'_0,v']]$ that reproduces $n'[\Psi'_0,v']$ starting from $\Phi'_0$.
	In terms of forces, $v'_\mathrm{Mxc}(\bz,t)$ is given by equating the equations of motion for $n'(\bz,t)$ for a non-interacting and interacting \labelcref{DressedEOM} system,
\begin{align*}
\nabla_\bz \cdot \left[ n'(\bz,t) \nabla_\bz v_\mathrm{Mxc}'(\bz,t) \right] &=
\nabla_\bz \cdot\left[ \mathbf{Q}'_\mathrm{kin,s}(\bz,t) - \mathbf{Q}'_\mathrm{kin}(\bz,t) \right] \nonumber \\
&- \nabla_\bz \cdot \mathbf{Q}'_\mathrm{int}(\bz,t) ,
\end{align*}
and models the force differences (not from the potentials) between the two systems to get the same forces in the two systems.
	Here $\mathbf{Q}'_\mathrm{kin,s}(\bz,t) = \imagi \braket<\Phi'(t)|[\hat{T}',\hat{\bj}'(\bz)]|\Phi'(t)>$ is the non-interacting dressed momentum-stress forces.

	The equation of motion for the dressed \ac{KS} density is,
\begin{align*}
\tfrac{\partial^2}{\partial t^2} n'(\bz,t) &=
\nabla_\bz \cdot \left\{ n'(\bz,t) \nabla_\bz [v'(\bz,t) + v'_\mathrm{Mxc}(\bz,t)] \right\} \\
&- \nabla_\bz \cdot \mathbf{Q}'_\mathrm{kin,KS}(\bz,t) .
\end{align*}
	Here $\mathbf{Q}'_\mathrm{kin,KS}(\bz,t) \!=\! \imagi \braket<\Phi'(t)|[\hat{T}',\hat{\bj}'(\bz)]|\Phi'(t)>$ is the dressed \ac{KS} momentum-stress forces, and equals $\mathbf{Q}'_\mathrm{kin,s}(\bz,t)$, but where we use the $_\mathrm{KS}$ to indicate it is of a \ac{KS} system.
	We mainly think of $v'_\mathrm{Mxc}(\bz,t)$ evaluated at self-consistency, as it usually is in time-dependent theory, but all equations of motion that we write up apply regardless.

	Like in \cref{sec:PhysicalDressedEquationsOfMotion}, we may again obtain the equation of motion for $n(\br,t)$ by integration.
	However, this time expressed in terms of quantities of the dressed \ac{KS} system.
	Since the results in \cref{sec:PhysicalDressedEquationsOfMotion} do not rely on specific properties of the subset of physical $v'(\br,t)$ and $w'(\bz,\bz')$, we may in fact even reuse the results from there for all but the $v'_\mathrm{Mxc}(\bz,t)$ term.
	The equation of motion that we get therefore also hold even for approximate $v'_\mathrm{Mxc}(\bz,t)$,
\begin{align*}
\!\!\! \tfrac{\partial^2}{\partial t^2} n(\br,t) \!&=\!
\nabla_\br \!\cdot\! \left[ n(\br,t) \nabla_\br v(\br,t) \!+\!\!\! \int \!\! \diff^M q \, n'(\bz,t) \nabla_\br v'_\mathrm{Mxc}(\bz,t) \!\right] \nonumber \\
&-\nabla_\br \cdot \left[ \bF_\mathrm{dip,KS}^\mathrm{d}(\br,t) \!+\! \bF_\mathrm{lin,KS}^\mathrm{d}(\br,t) \!+\! \mathbf{Q}_\mathrm{kin,KS}^\mathrm{d}(\br,t) \right] .
\end{align*}
	Here the dressed \ac{KS} stress and photon-matter forces are
\begin{align*}
\mathbf{Q}_\mathrm{kin,KS}^\mathrm{d}(\br,t) &= \imagi \braket<\Phi'(t)|[\hat{T},\hat{\bj}(\br)]|\Phi'(t)> , \\
\bF_\mathrm{dip,KS}^\mathrm{d}(\br,t) &= - \sum_{\alpha=1}^M \blambda_\alpha n(\br,t) (\blambda_\alpha \cdot \br) , \\
\bF_\mathrm{lin,KS}^\mathrm{d}(\br,t) &= \sum_{\alpha=1}^M \blambda_\alpha \int \diff^M q \tfrac{\omega_\alpha q_\alpha}{\sqrt{N}} n'(\bz,t) ,
\end{align*}
which only differ from the dressed physical $\mathbf{Q}_\mathrm{kin}^\mathrm{d}(\br,t)$, $\bF_\mathrm{dip}^\mathrm{d}(\br,t)$ and $\bF_\mathrm{lin}^\mathrm{d}(\br,t)$ in that they are based on $\Phi'(t)$.
	Especially, the two latter equal $\bF_\mathrm{dip}^\mathrm{d}(\br,t)$ and $\bF_\mathrm{lin}^\mathrm{d}(\br,t)$ up to how well $\Phi'(t)$ reproduces $n(\br,t)$ respectively $n'(\bz,t)$.
	If we also define $\bF_\mathrm{Mxc}(\br,t) \!=\! - \int \diff^M q n'(\bz,t) \nabla_\br v_\mathrm{Mxc}'(\bz,t)$, we find for the exact self-consistent \ac{KS} system that
\begin{align*}
\bF_\mathrm{Mxc}(\br,t) &=
\mathbf{Q}_\mathrm{kin}^\mathrm{d}(\br,t) - \mathbf{Q}_\mathrm{kin,KS}^\mathrm{d}(\br,t) + \mathbf{Q}_\mathrm{int}^\mathrm{d}(\br,t) \\
&= \mathbf{Q}_\mathrm{kin}(\br,t) - \mathbf{Q}_\mathrm{kin,KS}^\mathrm{d}(\br,t) \\
&+ \mathbf{Q}_\mathrm{int}(\br,t) + \tfrac{N-1}{N} \bF_\mathrm{lin}(\br,t) + \bF_\mathrm{dip}^{(2)}(\br,t) , \\
\bF_\mathrm{dip,KS}^\mathrm{d}(\br,t) &= \bF_\mathrm{dip}^\mathrm{d}(\br,t)
= \bF_\mathrm{dip}^{(1)}(\br,t) , \\
\bF_\mathrm{lin,KS}^\mathrm{d}(\br,t) &= \bF_\mathrm{lin}^\mathrm{d}(\br,t)
= \tfrac{1}{N} \bF_\mathrm{lin}(\br,t) ,
\end{align*}
so $\bF_\mathrm{Mxc}(\br,t) + \mathbf{Q}_\mathrm{kin,KS}^\mathrm{d}(\br,t) + \bF_\mathrm{dip,KS}^\mathrm{d}(\br,t) + \bF_\mathrm{lin,KS}^\mathrm{d}(\br,t) = \mathbf{Q}_\mathrm{kin}(\br,t) + \mathbf{Q}_\mathrm{int}(\br,t) + \bF_\mathrm{dip}(\br,t) + \bF_\mathrm{lin}(\br,t)$, and we get the same equation of motion for $n(\br,t)$ as in the physical system.
	We automatically get $\mathbf{Q}_\mathrm{kin,KS}^\mathrm{d}(\br,t) + \bF_\mathrm{dip}^{(1)}(\br,t) + \tfrac{1}{N}\bF_\mathrm{lin}(\br,t)$ (and the forces from $v(\br,t)$), while $\bF_\mathrm{Mxc}(\br,t)$ then has to model the rest, as discussed in the main text.

	Likewise we can again obtain the equation of motion for\!\! $p_\alpha(t)$, which also holds even for approximate $v'_\mathrm{Mxc}(\bz,t)$, by multiplying the equation of motion for $n'(\bz,t)$ by $\tfrac{q_\alpha}{\sqrt{N}}$, and integrating this over all $\bz$ coordinates (again we may reuse the results from \cref{sec:PhysicalDressedEquationsOfMotion}),
\begin{align*}
\tfrac{\partial^2}{\partial t^2} p_\alpha(t) &= - \omega_\alpha^2 p_\alpha(t) + \tfrac{\omega_\alpha}{N} \blambda_\alpha \cdot \mathbf{R}(t) - \tfrac{\dot{J}_\alpha(t)}{\omega_\alpha} \\
&- \tfrac{1}{\sqrt{N}} \int \diff\bz \, n'(\bz,t) \tfrac{\partial}{\partial q_\alpha} v'_\mathrm{Mxc}(\bz,t) .
\end{align*}
	To get the exact mode-resolved Maxwell equations the $v'_\mathrm{Mxc}(\bz,t)$ term has to contribute $\tfrac{N-1}{N} \omega_\alpha \blambda_\alpha \cdot \mathbf{R}(t)$.
	This is the contribution we get from $w'(\bz,\bz')$ in the interacting system.
	One may separate $-\tfrac{N-1}{N} \sum_{\alpha=1}^M \tfrac{\omega_\alpha q_\alpha}{\sqrt{N}} \blambda_\alpha \cdot \mathbf{R}(t)$ out of $v'_\mathrm{Mxc}(\bz,t)$ for this (which is a $\tfrac{N-1}{N}$ part of an identical term in the standard \ac{KS} mean-field approximation).

	To reproduce the exact $n'(\bz,t)$ we of course need to design $v'_\mathrm{Mxc}(\bz,t)$ to generate the right forces for $n'(\bz,t)$.
	However, more pragmatically we may also try to just get the right forces for $n(\br,t)$ only, or for $n(\br,t)$ and $p_\alpha(t)$, but we then have to compensate for the errors in $n'(\bz,t)$.
	This compensation is mainly needed to model $\bF_\mathrm{lin}(\br,t)$, as it is the one term in the equations of motion for $n(\br,t)$ and $p_\alpha(t)$ that strongly depends on the electron-photon correlations.
	This force can be highly sensitive though.
	This is a good reason to aim to get $n'(\bz,t)$ right.
	However, as only one $v'_\mathrm{Mxc}(\bz,t)$ get $n'(\bz,t)$ right, while many get $n(\br,t)$ and $p_\alpha(t)$ right, the latter may still be easier.

\section{Relationship with standard cavity quantum electrodynamics Kohn-Sham theory} \label{sec:RelationshipWithStandardKS}

	We consider the standard cavity \ac{QEKS} Hamiltonian \cite{tokatly2013}\!\!
\begin{align*}
\hat{H}_\mathrm{KS}(t) &= \sum_{k=1}^N \bigg\{
-\tfrac{1}{2} \nabla_{\br_k}^2 + v(\br_k,t) + v_\mathrm{mxc}(\br_k,t) \\
&\hspace{13mm} + \sum_{\alpha=1}^M [\blambda_\alpha \cdot \mathbf{R}(t) - \omega_\alpha p_\alpha(t)] \blambda_\alpha \cdot \br_k
\bigg\} \\
&+ \sum_{\alpha=1}^M \left[ - \tfrac{1}{2} \tfrac{\partial^2}{\partial p_\alpha^2} \!+\! \tfrac{1}{2}\omega_\alpha^2 p_\alpha^2 \!-\! \omega_\alpha p_\alpha \blambda_\alpha \!\cdot \mathbf{R}(t) \!+\! \tfrac{\dot{J}_\alpha(t)}{ \omega_\alpha} p_\alpha \right] ,
\end{align*}
of a non-interacting, non-correlated auxiliary \ac{KS} system that at self-consistency shares the same density $n(\br,t)$ and photon-displacements $p_\alpha(t)$ as the physical system.
	To ensure this, $\hat{H}_\mathrm{KS}(t)$ incorporates an explicit mean-field contribution, where $\mathbf{R}(t)$ and $p_\alpha(t)$ are the expectation values of $\hat{\mathbf{R}}$ and $p_\alpha$, and $v_\mathrm{mxc}(\br,t)$ denotes the remaining exchange-correlation potential.
	For the initial state $\Phi_0$, we in general use a linear combination of the eigenstates $\Phi_\mathrm{E} \prod_{\alpha=1\!}^M \varphi_{p_\alpha}(p_\alpha)$ of the \ac{KS} Hamiltonian $\hat{H}_\mathrm{KS}$ (especially, we generally use the ground state if $\Psi_0$ is a ground state).
	Here the electronic eigenstate $\Phi_\mathrm{E}$ is in general a weighted sum of Slater determinants $\Phi_{\br\sigma}$ of orbitals with spatial part $\varphi_{\br,k}(\br,t)$.
	This leads to the single particle equations
\begin{align*}
\imagi \tfrac{\partial}{\partial t} \varphi_{\br,k}(\br,t) &=
\bigg\{ -\tfrac{1}{2} \nabla_\br^2 + v(\br,t) + v_\mathrm{mxc}(\br,t) \\
&\hspace{5.5mm} + \!\sum_{\alpha=1}^M [\blambda_{\alpha\!} \!\cdot\! \mathbf{R}(t) \!-\! \omega_\alpha p_\alpha(t)] \blambda_{\alpha\!} \!\cdot \!\br \bigg\} \varphi_{\br,k}(\br,t) , \\
\imagi \tfrac{\partial}{\partial t} \varphi_{p_\alpha\!}(p_\alpha,t)
&= \Big[ - \tfrac{1}{2} \tfrac{\partial^2}{\partial p_\alpha^2} + \tfrac{1}{2} \omega_\alpha^2 p_\alpha^2 \\
&\hspace{6mm} - \omega_\alpha p_\alpha \blambda_\alpha \cdot \mathbf{R}(t) + \tfrac{\dot{J}_\alpha(t)}{\omega_\alpha} p_\alpha \Big] \varphi_{p_\alpha}(p_\alpha,t) ,
\end{align*}
and the force equations,
\begin{align*}
\tfrac{\partial^2}{\partial t^2} n(\br,t) &=
\nabla_\br \cdot \{n(\br,t) \nabla_\br [v(\br,t) + v_\mathrm{mxc}(\br,t)]\} \\
&- \nabla_\br \cdot \left\{ n(\br,t) \sum_{\alpha=1}^M \blambda_\alpha [\omega_\alpha p_\alpha(t) - \blambda_\alpha \cdot \mathbf{R}(t)] \right\} \\
&- \nabla_\br \cdot \mathbf{Q}_\mathrm{kin,KS}(\br,t), \\
\tfrac{\partial^2}{\partial t^2} p_\alpha(t) &=
- \omega_\alpha^2 p_\alpha(t) + \omega_\alpha \blambda_\alpha \cdot \mathbf{R}(t) - \tfrac{\dot{J}_\alpha(t)}{\omega_\alpha} ,
\end{align*}
where the prior includes the mean-field approximations to $\bF_\mathrm{lin}(\br,t)$ and $\bF_\mathrm{dip}(\br,t)$, and the second is the exact mode-resolved Maxwell equations due to the mean-field.
	Here $\mathbf{Q}_\mathrm{kin,KS}(\br,t) = \imagi \braket<\Phi(t)|[\hat{T},\hat{\bj}(\br)]|\Phi(t)>$ is the standard \ac{KS} momentum-stress forces.
	Note that one may solve the Maxwell equations analytically, $p_\alpha(t) = p_{\alpha,0} \cos(\omega_\alpha t) + \tfrac{\dot{p}_{\alpha,0}}{\omega_\alpha} \sin(\omega_\alpha t) + \int_0^t \sin[\omega_\alpha(t-t')] [\blambda_\alpha \cdot \mathbf{R}(t') - \tfrac{\dot{J}_\alpha(t')}{\omega_\alpha^2}] \diff t'$, and substitute the results into the single particle equations for $\varphi_{\br,k}(\br,t)$.
	This usually saves one to compute $\varphi_{p_\alpha}(p_\alpha,t)$.

	This problem can also be translated into a dressed \ac{KS} problem by exactly the same recipe as in the exact case.
	That is, we add $\hat{H}_\mathrm{aux}$ to $\hat{H}_\mathrm{KS}(t)$ and switch coordinates to $(q_{\alpha,1},...,q_{\alpha,N})$ for each mode to find $\hat{H}'_\mathrm{KS}(t) \!=\! \hat{T}' + \hat{V}'_\mathrm{KS}(t)$, with \scalebox{0.85}[1]{\!$v'_\mathrm{KS}(\bz,t) \!=\! v(\br,t) \!+\! v_\mathrm{mxc}(\br,t) \!+\! \sum_{\alpha=1}^M [\blambda_\alpha \!\cdot\! \mathbf{R}(t) \!-\! \omega_\alpha p_\alpha(t)] \blambda_\alpha \!\cdot\! \br$} \scalebox{0.87}[1]{$+ \sum_{\alpha=1}^M \left[ \tfrac{1}{2} \omega_\alpha^2 q_\alpha^2 \!-\! \tfrac{\omega_\alpha}{\sqrt{N}} q_\alpha \blambda_\alpha \cdot \mathbf{R}(t) \!+\! \tfrac{\dot{J}_\alpha(t)}{\sqrt{N} \omega_\alpha} q_\alpha \right]$}.
	We get no $\hat{W}'_\mathrm{KS}$, so this is a non-interacting dressed \ac{KS} Hamiltonian.
	Note that $v_\mathrm{mxc}(\br,t)$ depends only on $\br$, as usual in standard \ac{KS}, while the dressed $v'_\mathrm{Mxc}(\bz,t)$ depends on the full $\bz$.
	By noting that $\hat{H}_\mathrm{KS}(t) = \hat{H}_\mathrm{KS,E}(t) + \sum_{\alpha=1}^M \hat{H}_{p_\alpha}(p_\alpha-\bar{p}_\alpha(t))$ with $\bar{p}_\alpha(t) = - \tfrac{\dot{J}_\alpha(t)}{\omega_\alpha^3} + \tfrac{\blambda_\alpha \cdot \mathbf{R}(t)}{\omega_\alpha}$, we also see that $\hat{H}'_\mathrm{KS}(t) = \hat{H}_\mathrm{KS,E}(t) + \sum_{\alpha=1}^M \hat{H}'_{\mathrm{KS,P},\alpha}(t)$.
	Hence the eigenstates take the form discussed in \cref{sec:SymmetriesOfKohnShamSystems}.
	For the initial state we use $\Phi'_0 = \Phi_0 \chi_0$ ($\chi_0$ is usually the $q_{\alpha,k} \leftrightarrow q_{\alpha,l}$ symmetric ground state of $\hat{H}_\mathrm{aux}$, but may be any linear combination of eigenstates).\!
	This is a linear combination of eigenstates $\Phi_\mathrm{E} \prod_{\alpha=1}^M \varphi_{p_\alpha}(p_\alpha-\bar{p}_\alpha) \chi_\alpha$ of $\hat{H}'_\mathrm{KS} = \hat{H}_\mathrm{KS} + \hat{H}_\mathrm{aux}$ (usually just the ground state), and so one may also construct $\Phi'_0$ directly using $\hat{H}'_\mathrm{KS}$ instead of going through $\Phi_0$.
	Finally, we note that $\Phi'(t) = \Phi(t)\chi(t)$, and standard \ac{KS} therefore produces the $n'(\bz,t)$ of this $\Phi'(t)$.

	As a side remark, for the two electrons in a singlet state coupled with one mode examples of the main text, the standard mean-field and \ac{Mx} approximations to $v_\mathrm{mxc}(\br,t)$ are given by
\begin{align*}
v_\mathrm{MF}(\br,t) &= \tfrac{1}{2} \int \diff \br' \, n(\br',t) w(\br,\br') , \\
v_\mathrm{Mx}(\br,t) &= \tfrac{1}{2} \int \diff \br' \, n(\br',t) w(\br,\br') \\
&+ \tfrac{1}{2} (\blambda \cdot \br)^2 - \tfrac{1}{2} [\blambda \cdot \mathbf{R}(t)](\blambda \cdot \br) ,
\end{align*}
as our naming convention refers only to the light-matter interaction (while the Coulomb interaction in both cases is treated on an exact exchange level).
	The \ac{Mx} approximation again approximates the forces by their \ac{KS} values.

	In conclusion, standard cavity \ac{QEKS} is a special case of our dressed \ac{KS}, but where one reproduces the $n'(\bz,t)$ of the standard \ac{KS} system instead of the physical system (but the same $n(\br,t)$ and $p_\alpha(t)$).
	It is best implemented the standard way for numerical efficiency.
	The explicitly correlated orbitals of the dressed scheme have a limited extra numerical cost, but may capture electron-photon correlation better.

\section{Scaled Approximation} \label{sec:ScaledApproximation}

\def \ssMx /{$\sqrt{\text{s}}$\acs*{Mx}}
\def \sMx /{s\acs*{Mx}}
\def \tMx /{$\tilde{\text{\acs*{Mx}}}$}
\def \tssMx /{$\tilde{\sqrt{\text{s}}\text{\acs*{Mx}}}$}
\def \tsMx /{$\tilde{\text{s\acs*{Mx}}}$}

	In this \namecref{sec:ScaledApproximation} we discuss a first simple approximation for the \ac{Mxc} potential that makes use of the force relations that we established in \cref{sec:EOM}.
	Specifically, we redistribute the $\tfrac{N-1}{N} \bF_\mathrm{lin}(\br,t)$ forces originating from $w'(\bz,\bz')$ to $v'_\mathrm{KS}(\bz,t)$, by scaling by $N$ the term in $v'_\mathrm{KS}(\bz,t)$ responsible for $\tfrac{1}{N} \bF_\mathrm{lin}(\br,t)$.
	This simple idea ensures that the dressed \ac{KS} system trivially shares the exact same force expression for $\bF_\mathrm{lin}(\br,t)$ as the original system.


	For the dressed system,
\begin{align*}
v'(\bz,t) &= v(\br,t) + \sum_{\alpha=1}^M \Big[
\tfrac{1}{2} \omega_\alpha^2 q_\alpha^2 \underbrace{- \tfrac{\omega_\alpha}{\sqrt{N}} q_\alpha (\blambda_\alpha \cdot \br)}_{\hspace{-2cm}\rightarrow\, \bF_\mathrm{lin}^\mathrm{d} = \frac{1}{N}\bF_\mathrm{lin} \,\&\, \frac{1}{N} \omega_\alpha \blambda_\alpha \cdot \mathbf{R}(t)\hspace{-2cm}} \\
&+ \tfrac{1}{2} (\blambda_\alpha \cdot \br)^2 + \tfrac{\dot{J}_\alpha(t) q_\alpha}{\sqrt{N} \omega_\alpha} \Big] , \\
w'(\bz,\bz') &= w(\br,\br') + \sum_{\alpha=1}^M \Big[
(\blambda_\alpha \cdot \br)(\blambda_\alpha \cdot \br') \\
&\underbrace{- \tfrac{\omega_\alpha}{\sqrt{N}} q_\alpha (\blambda_\alpha \cdot \br')
- \tfrac{\omega_\alpha}{\sqrt{N}} q'_\alpha (\blambda_\alpha \cdot \br)}_{\hspace{-2cm}\rightarrow\, (N-1)\bF_\mathrm{lin}^\mathrm{d} = \frac{N-1}{N}\bF_\mathrm{lin} \,\&\, \frac{N-1}{N}\omega_\alpha \blambda_\alpha \cdot \mathbf{R}(t)\hspace{-2cm}} \Big] ,
\end{align*}
already a fraction $\tfrac{1}{N}$ of $\bF_\mathrm{lin}(\br,t)$, and of the $\omega_\alpha \blambda_\alpha \cdot \mathbf{R}(t)$ of each mode-resolved Maxwell equation, is generated by the dressed potential $v'(\bz,t)$, while the remaining parts $\tfrac{N-1}{N}$ are generated by $w'(\bz,\bz')$.
	In our dressed \ac{KS} scheme those interactions have to be mimicked by the mean-field exchange correlation potential $v'_\mathrm{Mxc}(\bz,t)$.


	An interesting idea to do this is to first redistribute the $\tfrac{N-1}{N} \bF_\mathrm{lin}(\br,t)$ forces originating from $w'(\bz,\bz')$ to $v'(\bz,t)$ by scaling the corresponding part $- \tfrac{\omega_\alpha}{\sqrt{N}} q_\alpha (\blambda_\alpha \cdot \br)$ of $v'(\bz,t)$ by $N$.
	That is, we consider a $\tilde{H}'(t)$ system with,
\begin{align*}
\tilde{v}'(\bz,t) &= v(\br,t) + \sum_{\alpha=1}^M \Big[
\tfrac{1}{2} \omega_\alpha^2 q_\alpha^2 \underbrace{- \sqrt{N} \omega_\alpha q_\alpha (\blambda_\alpha \cdot \br)}_{\hspace{-2cm}\rightarrow\, \bF_\mathrm{lin} \,\&\, \omega_\alpha \blambda_\alpha \cdot \mathbf{R}(t)\hspace{-2cm}} \\
&+ \tfrac{1}{2} (\blambda_\alpha \cdot \br)^2 + \tfrac{\dot{J}_\alpha(t) q_\alpha}{\sqrt{N} \omega_\alpha} \Big] + \tilde{v}'_\mathrm{Mxc}(\bz,t) , \\
\tilde{w}'(\bz,\bz') &= w(\br,\br') + \sum_{\alpha=1}^M (\blambda_\alpha \cdot \br)(\blambda_\alpha \cdot \br') .
\end{align*}
	Even for $\tilde{v}'_\mathrm{Mxc}(\bz,t) = 0$, this system shares the same total force expression for $n(\br,t)$ as the original physical system with the minor redistribution that
\begin{align*}
\tilde{\mathbf{Q}}_\mathrm{int}^\mathrm{d}(\br,t) &=
-2 \int \diff \br' \rho_2(\br,\br',t) \nabla_\br w(\br,\br') \\
&-2 \sum_{\alpha=1}^M \blambda_\alpha \int \diff \br' \rho_2(\br,\br',t) (\blambda_\alpha \cdot \br') , \\
\tilde{\bF}_\mathrm{lin}^\mathrm{d}(\br,t) &= \underbrace{N}_{!} \sum_{\alpha=1}^M \blambda_\alpha \int \diff^M q \tfrac{\omega_\alpha q_\alpha}{\sqrt{N}} n'(\bz,t) .
\end{align*}
	Also the mode-resolved Maxwell equations stay identical
\begin{align*}
\partial_t^2 p_\alpha(t) = - \omega_\alpha^2 p_\alpha(t) + \omega_\alpha \blambda_\alpha \cdot \mathbf{R}(t) - \tfrac{\dot{J}_\alpha(t)}{\omega_\alpha} ,
\end{align*}
with the only redistribution that now all of $\omega_\alpha \blambda_\alpha \cdot \mathbf{R}(t)$ originates from $\tilde{v}'(\bz,t)$ instead of $\tfrac{1}{N}$ from $v'(\bz,t)$ and $\tfrac{N-1}{N}$ from $w'(\bz,\bz')$.
	That is, the full RHS now originates from $\tilde{v}'(\bz,t)$ alone.
	To reproduce $n'(\bz,\!t)$ we still need the right $\tilde{v}'_\mathrm{Mxc}(\bz,t)$ though (i.e., $\tilde{H}'(t)$ is a \ac{KS} system).
	Without it, the underlying system solved $\tilde{H}'(t)$, which enacts forces $\tilde{\bF}'(\bz,t)$ on the dressed density $n'(\bz,t)$, will still exhibit deviating forces from the original system $\hat{H}'(t)$.
	That is, $\tilde{\Psi}'(t)$ still differs from $\Psi'(t)$, leading to different forces even for the physical observables $n(\br,\!t)$ and $p_\alpha(t)$ despite the force expressions are now exact.%
\footnote{%
	$\tilde{H}'(t)$ also breaks all the physical symmetries, however, the $n(\br,t)$ and $p_\alpha(t)$ force expressions still apply as is, as the one expression that relied on these properties for $\hat{W}'$ is no longer present for $\tilde{W}'$.
}
	This is especially so as this idea has not been designed with $n'(\bz,t)$ in mind.
	Indeed, $\bF_\mathrm{lin}(\br,t)$ can be very sensitive to this, since it depends strongly on the electron-photon correlations in $n'(\bz,t)$, which are treated differently in the $\tilde{H}'(t)$ system.
	Hence we cannot expect $\tilde{H}'(t)$ to perform well in general without $\tilde{v}'_\mathrm{Mxc}(\bz,t)$.
	However, it can still be very precise for specific domains of applicability.


	In practise, this all leads us to rewrite the $v'_\mathrm{KS}(\bz,t) = v'(\bz,t) + v'_\mathrm{Mxc}(\bz,t)$ of the usual dressed \ac{KS} system as
\begin{align*}
v'_\mathrm{KS}(\bz,t) = \bar{v}'(\bz,t) + v'_\mathrm{sMxc}(\bz,t) ,
\end{align*}
where $\bar{v}'(\bz,t)$ is shorthand for $\tilde{v}'(\bz,t)$ without $\tilde{v}'_\mathrm{Mxc}(\bz,t)$.
	This generates the exact force expression for $\bF_\mathrm{lin}(\br,t)$ and for the full mode resolved Maxwell equations even for $v'_\mathrm{sMxc}(\bz,t) = 0$, while $v'_\mathrm{sMxc}(\bz,t) = v'_\mathrm{Mxc}(\bz,t) - \tfrac{1-N}{\sqrt{N}} \omega_\alpha q_\alpha (\blambda_\alpha \!\cdot\! \br)$ represents the remaining \ac{Mxc} potential after scaling (i.e., it generates the other forces and further corrections to the bilinear forces, e.g., $\tilde{v}'_\mathrm{Mxc}(\bz,t)$).
	If we neglect $\tilde{v}'_\mathrm{Mxc\!}(\bz,_{\!}t)$, and approximate the contribution from $\tilde{w}(\bz,\bz')$ by its \ac{KS} value, we get the \sMx/ approximation,
\begin{align*}
\nabla_\bz \cdot \left[ n'(\bz,t) \nabla_\bz v'_\mathrm{sMx}(\bz,t) \right] = - \nabla_\bz \cdot \imagi \braket<\Phi'(t)|[\tilde{W}',\hat{\bj}'(\bz)]|\Phi'(t)> .
\end{align*}
	This is the scaled analogue of the \ac{Mx} approximation, and differs only in the bilinear forces here are treated by redistribution instead of using the \ac{KS} values.


	We now study how the \sMx/ approximation performs for the two examples in the main text (compared with the \ac{Mx} results presented there).
	Note that in both cases, i.e., for two particles in a singlet state coupled to one mode with no external current $J(t)=0$ the \sMx/ approximation reduces to
\begin{align*}
\bar{v}'(\br,q,t) &= v(\br,t) + \tfrac{1}{2} \omega^2 q^2 - \sqrt{2} \omega q (\blambda \cdot \br) + \tfrac{1}{2} (\blambda \cdot \br)^2 , \\
v'_\mathrm{sMx}(\br,q,t) &= \tfrac{1}{2} \int \diff^3 \br' \, n(\br',t) w(\br,\br') + \tfrac{1}{2} [\blambda \cdot \mathbf{R}(t)](\blambda \cdot \br) .
\end{align*}
	Also note that $v'_\mathrm{sMx}(\bz,t)$ is simply $v'_\mathrm{Mx}(\bz,t)$ without the two bilinear coupling terms.

\begin{figure} [H]
\includegraphics[width=8.6cm]{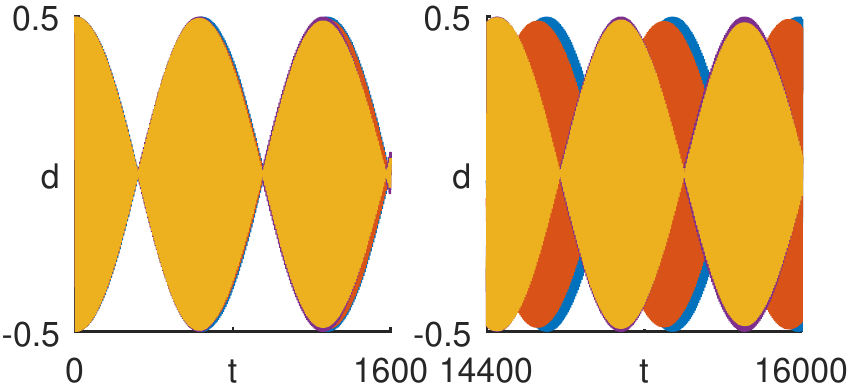}
\caption{(color online) The exact (orange), dressed \sMx/ (lilac), dressed \ac{Mx} (red) and standard \ac{Mx} (blue) dipole moment of two electrons on two sites coupled to one mode.}
\label{fig:TwoSiteScaled}
\end{figure}

	We first compare our new \sMx/ results with the previously presented \ac{Mx} results for the two-site case in \cref{fig:TwoSiteScaled}.
	Dressed \sMx/ (lilac) improves the accuracy for the given time-period to almost perfect agreement, and greatly outperforms both dressed \ac{Mx} (red) and standard \ac{Mx} (blue).
	However, we should note that if one for example increases $\lambda$ to $0.5$ (not shown), where we only looked at $t = [0,100]$, dressed \ac{Mx} actually slightly outperforms dressed \sMx/, but both still perform very decent.

\begin{figure} [H]
\includegraphics[width=8.6cm]{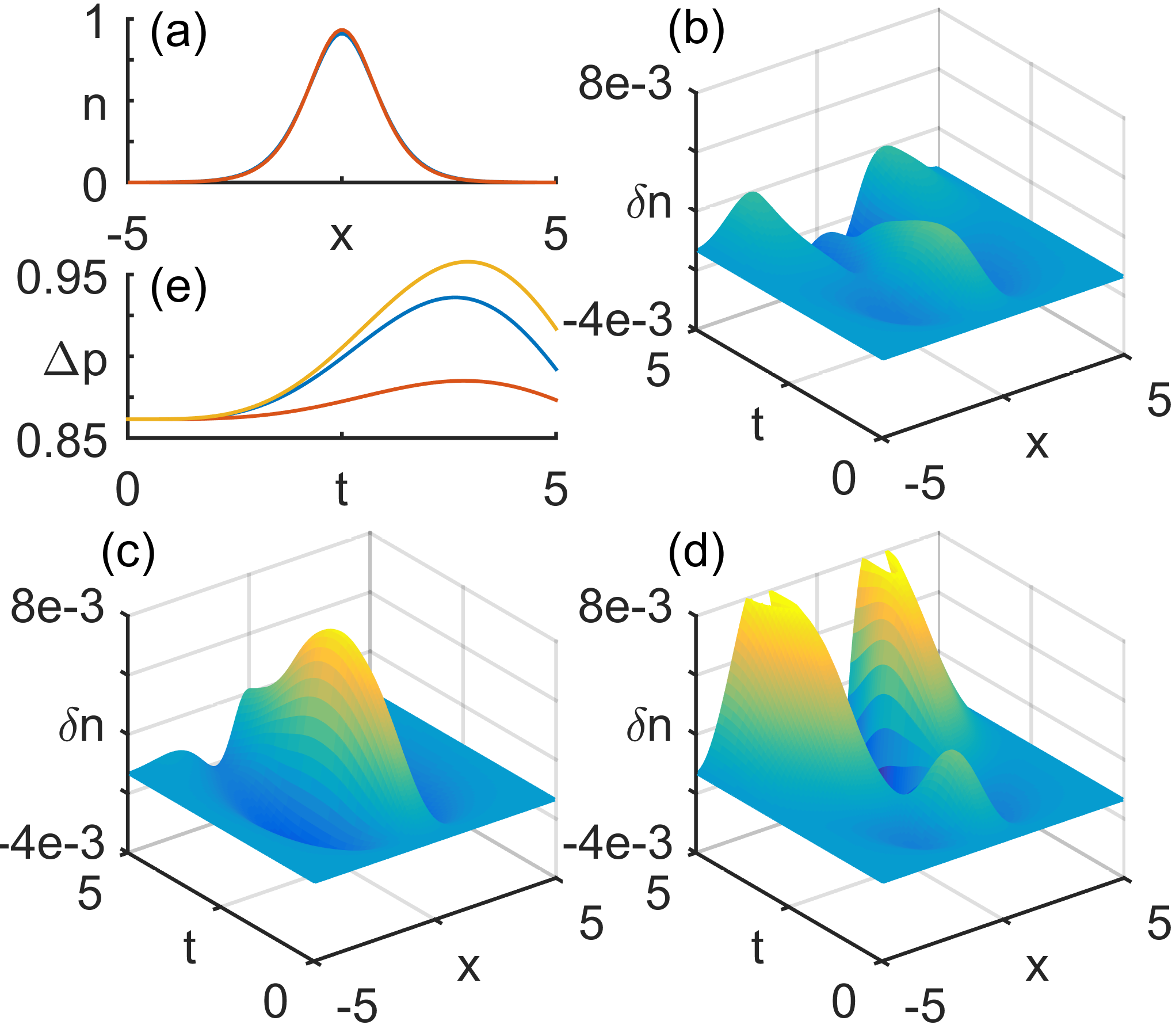}
\caption{(color online) (a) The exact (blue) and \ac{Mx} (red) ground-state densities of the bare ($\lambda=0$) one-dimensional helium model. The corresponding changes in the (b) exact (c) dressed \ac{Mx} and (d) dressed \sMx/ density, $\delta n(x,t)= n(x,t) - n_0(x)$, when placed inside a cavity, and (e) the exact (blue), dressed \ac{Mx} (red) and dressed \sMx/ (orange) field fluctuations.}
\label{fig:HeliumScaled}
\end{figure}

	Let us now address the example of spontaneous emission for the one-dimensional helium model in \cref{fig:HeliumScaled}.
	Dressed \sMx/ (orange and (d)) improves the photon-field fluctuations, and the density is better at short times but worsens for longer times compared to dressed \ac{Mx} (red and (c)).
	The density improvement for short times relies on error cancellation though.
	That is, we only have a density change due to coupling with the photon mode, so the change is mainly due to $\bF_\mathrm{lin}(\br,t)$ and $\bF_\mathrm{dip}(\br,t)$.
	The errors in these are in opposite directions, so while the $\bF_\mathrm{lin}(\br,t)$ errors are here larger for dressed \sMx/ than dressed \ac{Mx}, there is still an improvement for short times where the $\bF_\mathrm{dip}(\br,t)$ errors dominate.
	To understand this case better, an analysis of the individual contributions, and their approximations (and how they work together), is required.
	We focus in the following on the bilinear coupling, since essentially only this part of $\hat{H}(t)$ is treated differently in dressed \ac{KS} than standard \ac{KS}, and it is also the difference between our dressed \sMx/ and dressed \ac{Mx}.
	Here the $\tilde{H}'(t)$ system is very useful, since it lets us study the effect of approximating the bilinear coupling alone (as the other parts are the exact ones).
	In the following, we therefore compare exact results (obtained using $\hat{H}(t)$) with \tsMx/ (using $\tilde{H}'(t)$ with $\tilde{v}'_\mathrm{Mxc}(\bz,t)=0$) and \tMx/ (using $\tilde{H}'(t)_{\!}$ but with the \ac{Mx} approximation for the bilinear part instead of the scaled potential, i.e., with $- \sqrt{2} \omega q (\blambda \cdot \br) + \tilde{v}'_\mathrm{Mxc}(\bz,t) \!=\! - \tfrac{\omega}{\sqrt{2}} q (\blambda \cdot \br) - \tfrac{\omega}{2} p(t) (\blambda \cdot \br) - \tfrac{\omega}{2\sqrt{2}} q \blambda \cdot \mathbf{R}(t)$, where the $v'_\mathrm{Mx}(\bz,t)$ part here fails to contribute by symmetry, i.e., $p(t)=\mathbf{R}(t)=0$).
	For convenience we further change to ground state \ac{KS}, as the physics remains the same.

\begin{figure} [H]
\includegraphics[width=8.6cm]{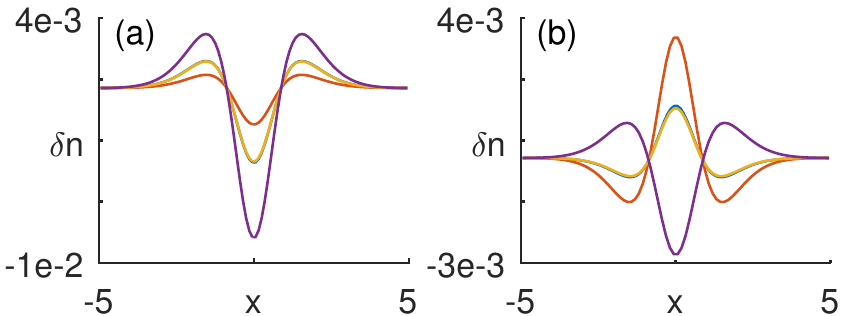}
\caption{(color online) The changes in the exact (blue, hidden behind orange), \tMx/ (red), \tssMx/ (orange) and \tsMx/ (lilac) ground state density, $\delta n(x) = n_\lambda(x) - n_{\lambda=0}(x)$, of a one-dimensional helium model (a) without $w(x,x')$ and $\lambda^2$ terms and (b) without $w(x,x')$.}
\label{fig:Bilinear}
\end{figure}

	We start by studying the effect of the bilinear coupling alone.
	That is, we set $w(\br,\br')$ and the quadratic coupling (all $\lambda^2$ terms) to zero for both the exact and $\tilde{H}'$ systems.
	This leads to \cref{fig:Bilinear} (a), where the shown change in the ground state density is due to the bilinear coupling alone.
	Here \tsMx/ (lilac) gives twice the exact density change (blue, hidden behind orange), while \tMx/ (red) only gives half (as the \ac{Mx} potential fails to contribute).
	Hence we need further corrections $\tilde{v}'_\mathrm{Mxc}(\bz)$.
	We can reproduce the density very well by scaling with $\sqrt{N}$ instead of $N$ though (i.e., using $-\omega q(\blambda \cdot \br)$ instead of $-\sqrt{2}\omega q(\blambda \cdot \br)$ in $\tilde{v}'(\bz)$).
	We call this \tssMx/ (orange), which may have a domain of applicability to the extent it can be generalised to more electrons and with a further correction asymmetric cases.
	The change in the exact field fluctuations $0.0102$ ($0.7091$ minus $0.6989$ at $\lambda=0$) is reproduced very accurately by \tsMx/ $0.0103$, while \tMx/ only yields a quarter of the change $0.0025$, and \tssMx/ half the change $0.0050$ (all vs $0.6989$).	
	So with scaling alone there is in this case a trade-off where we can either get the density (\tssMx/) or field fluctuations (\tsMx/) right, but not both, this requires some $\tilde{v}'_\mathrm{Mxc}(\bz)$.

	If we include the $\lambda^2$ terms again we get \cref{fig:Bilinear} (b).
	Here (at least for this coupling strength) the density changes almost equal the sum of the density change for quadratic alone (not shown), and for bilinear alone.
	Hence we can in this case consider these two approximations separately, and \tssMx/ together with a good approximation for the quadratic terms will yield excellent densities.
	Note how the \tsMx/ density change is opposite of what it should be; it is essential to get the right scale to balance the effect of the bilinear and quadratic couplings against each other.
	The field fluctuations are barely changed by including the $\lambda^2$ terms (all changes are reduced but at most by $0.0004$), so they are still reproduced very accurately by \tsMx/.

\begin{figure} [H]
\includegraphics[width=8.6cm]{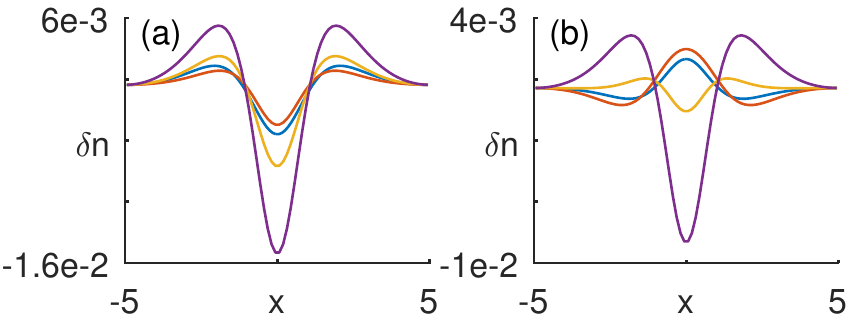}
\caption{(color online) The changes in the exact (blue), \tMx/ (red), \tssMx/ (orange) and \tsMx/ (lilac) ground state density, $\delta n(x) = n_\lambda(x) - n_{\lambda=0}(x)$, of a one-dimensional helium model (a) without $\lambda^2$ terms and (b) with all terms.}
\label{fig:BilinearAndInteraction}
\end{figure}

	Including $w(\br,\br')$ again (but again no $\lambda^2$ terms at first) we arrive at \cref{fig:BilinearAndInteraction} (a), where the density change is again only due to the bilinear coupling.
	Here regular \tMx/ (red) outperforms \tssMx/ (orange), which now gives a too strong change, and \tsMx/ (lilac), which is now even more too strong.
	This effect of $w(\br,\br')$ on the effect of the bilinear coupling limits how good results we can expect using any fixed scaling on its own (e.g., \ssMx/ or \sMx/).
	We need some $\tilde{v}'_\mathrm{Mxc}(\bz)$.
	Especially, if we include the $\lambda^2$ terms again in \cref{fig:BilinearAndInteraction} (b), it again just adds its own change without affecting the other changes much but now only \tMx/ yields a qualitatively right density change, clearly showing the issue in using wrong scaling.
	\Cref{fig:BilinearAndInteraction} (a) is also a clear example of that while $v'(\bz)$ gives us $\tfrac{1}{N}$ of the force expression for $\bF_\mathrm{lin}(\br)$, it only gives us $\tfrac{1}{N}$ of the force if we use the exact $v'_\mathrm{Mxc}(\bz)$.
	In this case we then just happen to get more of the force, so \tMx/ (for which the only contribution to $\bF_\mathrm{lin}(\br)$ in this case is that from $v'(\bz)$) is closer to the exact than one would expect.
	The change in the exact field fluctuations $0.0183$ ($0.8797$ minus $0.8615$ for $\lambda\!=\!0$) is now also in between \tsMx/ $0.0255$ and \tssMx/ $0.0121$, so this picture is also less clear here (\tMx/ gives $0.0059$, all vs $0.8615$).
	The field fluctuations are still barely changed by including the $\lambda^2$ terms though (all changes are again reduced and at most by $0.0012$).
	
\begin{figure} [H]
\includegraphics[width=8.6cm]{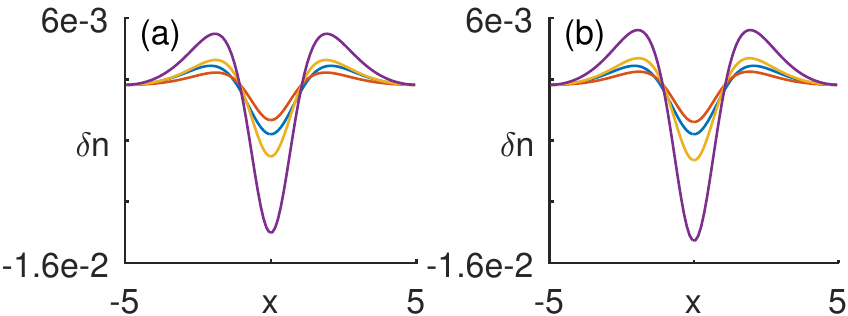}
\caption{(color online) (a) The changes in the exact (blue), dressed \ac{Mx} (red), dressed \ssMx/ (orange) and dressed \sMx/ (lilac) ground state density, $\delta n(x) = n_\lambda(x) - n_{\lambda=0}(x)$, of a one-dimensional helium model without $\lambda^2$ and (b) using an improved approximation for the interaction $w(x,x')$.}
\label{fig:BilinearAndInteractionKS}
\end{figure}

	Finally, we show that we get the same density changes from the bilinear coupling when using \ac{KS} also for $w(\br,\br')$,\!\! i.e., using $\hat{H}'_\mathrm{KS}$ instead of $\tilde{H}'$, though it treats the electron correlations differently.
	Using exact exchange for $w(\br,\br')$, the bare \ac{KS} and exact ground state differ.
	However, the density change compared to the bare density is still only due to the bilinear coupling (as we include no $\lambda^2$ terms), and in \cref{fig:BilinearAndInteractionKS} (a) we confirm that the dressed \ac{Mx}, \ssMx/ and \sMx/ density changes are nearly the same as using $\tilde{H}'$ (\cref{fig:BilinearAndInteraction} (a)).
	This remains true also using more precise approximations for $w(\br,\br')$ such as $v'_\mathrm{KS}(\bz) = \bar{v}'(\bz) + v'_\mathrm{sMxc,bare}(\bz) + \tfrac{1}{2} \int \diff^3 \br' \, [n(\br')-n_\mathrm{bare}(\br')] w(\br,\br')$, which uses the exact bare s\ac{Mxc} potential (so the bare densities now agree) added the exact exchange backreaction to the density change.
	The corresponding density changes in \cref{fig:BilinearAndInteractionKS} (b) are even closer to \cref{fig:BilinearAndInteraction} (a) than in \cref{fig:BilinearAndInteractionKS} (a).


	In conclusion, we cannot expect to find a scaling that works well on its own in general, especially given the dependence on $w(\br,\br')$, without further modifications.
	However, scaling may still be very accurate in special cases, as seen in our two-site case for the dipole moment, or field fluctuations of our one-dimensional helium model.
	Whether the \ssMx/ approximation can be generalised to more electrons, and if it has an area of applicability when $w(\br,\br')=0$, for similar cases, remains to be seen.
	Clearly there is still much to investigate about scaling in general, for example also by studying the $n'(\bz,t)$ equation of motion.

\section{Outlook}

	In this supplemental material we have provided many details of the properties of the dressed physical and \ac{KS} systems.
	We have further highlighted how the dressed \ac{KS} construction reproduces the exact dressed density $n'(\bz,t)$ by emulating the missing forces of the dressed physical system and thereby that of the original physical system.
	Since the dressed \ac{KS} system is based on explicitly correlated orbitals, approximations to the \ac{Mxc} potential are automatically correlated (at the expense of violating certain physical symmetries).
	However, also the standard \ac{QEKS} construction can be recast in the dressed picture.
	This allows one to use standard functionals and investigate them in the dressed setting.
	This interesting option together with the scaled approximation and all the relations between the physical, dressed and \ac{KS} forces is the subject of ongoing work to find more accurate functionals for dressed \ac{QEKS}, standard \ac{QEKS} as well as quantum-electrodynamical reduced density matrix functional theory \cite{buchholz2018reduced}.


\bibliographystyle{apsrev4-1}
\bibliography{Library}
